\documentstyle[12pt,amssymb]{article}

\begin{document}
\author{
D.~Costa$^\dag$, F.~Micali$^\dag$, F. Saija$^\ddag$, and
P.~V.~Giaquinta$^\dag$\footnote{Corresponding author; E-mail:
{\tt Paolo.Giaquinta@unime.it}} \\ \\ 
{\small {\it \dag~Istituto Nazionale per la Fisica della Materia (INFM) and}}\\
{\small {\it Universit\`a degli Studi di Messina, Dipartimento di Fisica}} \\
{\small {\it Contrada Papardo, C.P. 50 - 98166 Messina, Italy}}  \\
{\small {\it \ddag~CNR - Istituto per i Processi Chimico-Fisici, sez. Messina}} \\
{\small {\it Via La Farina 237 -  98123 Messina, Italy}} } 
\title{Entropy and correlations in a fluid of hard spherocylinders:
The onset of nematic and smectic order}
\date{}
\maketitle

\begin{abstract}
Hard spherocylinders (cylinders of length $L$ and diameter $D$
capped at both ends with two hemispheres) provide a suitable model
for investigating entropy-driven, mesophase formations in real
colloidal fluids that are composed of rigid rodlike molecules. We
performed extensive Monte Carlo simulations of this model fluid
for elongations in the range $3 \leq L/D \leq 5$ and up to $L/D =
20$, in order to investigate the relative importance of
translational and orientational correlations allowing for the
emergence of nematic or smectic order in the framework of the
so-called residual multi-particle entropy (RMPE). The vanishing of
this quantity, which includes the re-summed contributions of all
spatial correlations involving more than two particles, signals
the structural changes which take place, at increasing densities,
in the isotropic fluid. We found that the ordering thresholds
detected through the zero-RMPE condition systematically correlate
with the corresponding phase-transition points, whatever the
nature of the higher-density phase coexisting with the isotropic
fluid.

\end{abstract}


\section{Introduction}
A systematic approach to the study of the relation between
configurational entropy, spatial correlations and local ordering
in a fluid of elongated particles was recently outlined
in~\cite{csg}, where the Authors used the theoretical framework
that had been originally introduced by Giaquinta and Giunta in
order to identify the ``hidden'' signature of freezing in the
structural properties of a homogeneous and isotropic fluid of hard
spheres \cite{gg}. Their approach was based on the well known
multi-particle correlation expansion of the statistical entropy,
first established for a finite system by H.~S.~Green \cite{green}
and later generalized to an infinite open system by
R.~E.~Nettleton and M.~S.~Green~\cite{ng}: \begin{equation}\label{s:sum} s_{\rm
ex}=\sum_{n = 2}^{\infty} s_{n}\ , \end{equation} where $s_{\rm ex}$ is the
excess entropy per particle in units of the Boltzmann constant and
the ``$n$-body entropies'' ($s_{n}$) are obtained from the
integrated contribution of spatial correlations between $n$-tuples
of particles. Note that the one-body term coincides with the
entropy of the corresponding ideal gas. The pair entropy of a
homogeneous and isotropic fluid of {\it nonspherical}
molecules can be written as \cite{lapa}: \\
\begin{equation}\label{eq:s2mol}
s_{2}(\rho) =  -\frac{1}{2}
\frac{\rho}{\Omega^2} \int \{ g(r,\omega^2) \ln [g(r,\omega^2)] 
- g(r,\omega^2) + 1\} {\rm d} {\bf r} {\rm d} \omega^2 \,,
\end{equation} 
where $g(r,\omega^2)$ is the pair distribution function (PDF)
which depends on the relative separation $r$ between a pair of
molecules and on the set of Euler angles $\omega^2
\equiv[\omega_{1},\omega_{2}]$ that specify the absolute
orientations of the two particles in the laboratory reference
frame. The quantity $\Omega$ represents the integral over the
Euler angles of one molecule ($\Omega=4\pi$ for linear molecules),
while $\rho$ is the particle number density.

The approach originally set up by Giaquinta and Giunta focused on
the properties of the so-called residual multi-particle entropy
(RMPE): \begin{equation}\label{s:sex} \Delta s \equiv s_{\rm ex} - s_{2}\ . \end{equation}
The quantity $\Delta s$ includes, by definition, the re-summed
contributions of all correlations involving more than two
particles~\cite{gg}. This quantity, despite its minor quantitative
relevance in the overall entropy balance, is far from being
irrelevant in that it conceals significant indications on the
statistical thermodynamics of the fluid that are intimately
related to the role played by high-order density correlations in
the system. The noticeable feature exhibited by the RMPE, as
compared to the pair entropy (a negative-definite quantity), is
its non-monotonic behavior. In fact, $\Delta s\left( \rho \right)$
starts decreasing for increasing densities until the function
attains a minimum beyond which it undergoes a sharp increase up to
positive values. This behavior may come out, at first glance,
somewhat unexpected in that one would probably surmise that, in
analogy with the monotonic behavior of the two-body term $s_2$,
the growth of higher-order density correlations that is induced by
a further compression of the system should throughout contribute
to a contraction of the configurational states that are accessible
to the system for smaller and smaller volumes, as is actually the
case from low to intermediate densities. Instead, one observes a
``slowing down'', for increasing pressures, in the reduction rate
of the average number of states with respect to the
non-interacting system in an equivalent thermodynamic condition:
in fact, the positive cumulative contribution of multiparticle
correlations to the entropic balance moderately damps the
systematic drop that is caused by the dominant pair term. This
effect is suggestive of the emergence of a new structural
condition in the system. More specifically, it indicates that
multiparticle correlations have started to build up some type of
ordering, on a local scale, as a result of compelling entropic
constraints. The associated signature is the vanishing of the
RMPE: {\it a posteriori}, it is found to herald in a very
sensitive way the occurrence - in a proximate range of densities
and/or temperatures - of a phase transition of the fully
disordered fluid into a more structured phase. By now, the
reliability of such a one-phase ordering criterion has been
extensively documented for a variety of model
fluids~\cite{spg1,sg}, also in two dimensions \cite{spg2}.

With specific reference to a fluid of hard spherocylinders, i.e.,
cylinders of length $L$ and diameter $D$ capped at both ends with
two hemispheres, preliminary evidence was reported in~\cite{csg}
on the RMPE indication concerning the isotropic-nematic (I-N)
transition. In the case of simple fluids composed of spherical
particles a number of one-phase criteria have been introduced in
the past with the aim of locating the transition from the liquid
to the solid phase without resorting to the knowledge of the free
energy of the two coexisting phases~\cite{monson}. However, no
such simple criteria are available for nematic fluids. In this
respect, the zero-RMPE criterion stands as, perhaps, a unique
exception in that it can be extended in a straightforward way to
fluids composed of elongated particles to deal with the transition
from the homogeneous and isotropic fluid to a liquid-crystalline
phase.

Since the seminal work of Onsager~\cite{ons}, the model of hard
spherocylinders has been diffusely investigated, notwithstanding
its simplicity, in order to study excluded-volume effects in real
liquid crystals. In fact, hard spherocylinders exhibit --
depending on the value assumed by the aspect ratio $L/D$ -- a
surprising variety of
mesophases~\cite{frenkel1,flb,verfre,jawil,bofr}. In~\cite{csg}
constant-pressure Monte Carlo simulations were carried out for
$L/D=5$ only, a case where the disordered fluid undergoes, under
compression, a first-order transition to a partially ordered
nematic phase. In this paper we have extended the above
calculations to lower values of the aspect ratio in the range $3
\leq L/D \leq 5$, thus encompassing a region where the homogeneous
and isotropic fluid undergoes a transition to a fully ordered
solid (S) phase as well as to the smectic-A (SmA) phase, a layered
orientationally ordered phase characterized by a one-dimensional
density modulation along the direction $\bf n$ of the molecular
alignment~\cite{bofr}. We have carried out simulations also in the
limit of very elongated particles, specifically for $L/D=20$, in
order to analyze the behavior of the RMPE in a regime where the
I-N transition point can be determined, asymptotically, in an
exact way~\cite{ons,lek,wj}. This is also the range of aspect
ratios relevant for the modelling of monodisperse aqueous
suspensions of rodlike particles like the {\it tobacco mosaic
virus}, a plant virus which shows an isotropic-nematic transition
as well as a smectic phase at higher concentrations~\cite{uist}.

The paper is organized as follows: The theoretical framework and
the numerical simulation technique, as applied to the case of
cylindrical molecules with head-tail symmetry, are described in
Sects.~2~and~3, respectively. The results are presented in
Sect.~4 while Sect.~5 is devoted to concluding remarks.

\section{Theory}
In order to evaluate the RMPE from Eq.~(\ref{s:sex}), one needs to
calculate the excess and pair entropies of the fluid. The former
quantity can be obtained by integrating the equation of state
(EoS) along a thermodynamic path: \begin{equation}\label{s:free_ex} s_{\rm ex}
(\rho) = s_{\rm ex} (\bar{\rho}) - \int_{\bar{\rho}}^{\rho} \left[
\frac{{\beta} P} {\rho'} - 1 \right] \frac{{\rm d}{\rho'}}{\rho'}
\,. \end{equation} In Eq.~(\ref{s:free_ex}) $P$ is the pressure, $\beta$ is
the inverse temperature in units of the Boltzmann constant, and
$\bar{\rho}$ is the density of a suitably chosen reference state.
The EoS of the fluid was obtained through a Monte Carlo sampling
of the system at different pressures, while the quantity $s_{\rm
ex}(\bar{\rho})$ was calculated using the Widom test-particle
insertion method~\cite{widom,smfr} at low enough densities.

As for the pair entropy, this quantity was obtained, according to
Eq.~(\ref{eq:s2mol}), after integration of the PDF which, in turn,
was decoupled in the form~\cite{laka}: \begin{equation}\label{gtot}
g(r,\omega^2)=g(r)g(\omega^2|r) \;, \end{equation} where $g(r)$ is the radial
distribution function of the molecular centers of mass that is
obtained after integrating the full PDF over the set of Euler
angles, while $g(\omega^2|r)$ represents the conditional
distribution function, normalized to $\Omega^2$, for the relative
orientation of a pair of molecules whose centers of mass lie at a
distance $r$. Using Eq.~(\ref{gtot}), the pair entropy was
separated into the sum of a translational ($s_2^{\rm tr}$) and an
orientational part ($s_2^{\rm or}$), after sorting out an
additional excluded-volume contribution which arises in
Eq.~(\ref{eq:s2mol}) from the integration over space regions where
$g(r,\omega^2)=0$:

\begin{equation}\label{s:s2} s_{2}=- B_2\rho + s_{2}^{\rm tr} + s_{2}^{\rm or}
\;. \end{equation}

\noindent In Eq.~(\ref{s:s2}), $B_2$ is the second virial
coefficient of hard spherocylinders: \begin{equation}\label{b2} B_{2} =
\frac{2}{3}\pi D^3 + \pi D^2 L + \frac{\pi}{4} D L^2 \;. \end{equation} The
translational contribution is formally analogous to the pair
entropy of a simple fluid with no rotational degrees of freedom:
\begin{equation}\label{s2gl} s_{2}^{\rm tr} = -\frac{1}{2} \rho \int [ g(r) \ln
g(r) - g(r) + 1 ] {\rm d}{\bf r} \,, \end{equation} where the integration is
carried out over non-overlapping regions. As for the orientational
term, one finds: \begin{equation}\label{s:s2or} s_{2}^{\rm or} = \rho \int
g(r)S^{\rm or}(r) {\rm d}{\bf r} \,, \end{equation} where \begin{equation}\label{s:s2orbig}
S^{\rm or}(r) = -\frac{1}{2\Omega^{2}} \int g(\omega^2|r) \ln
[g(\omega^2|r)] {\rm d} \omega^2 \;. \end{equation} Even for uniaxial
molecules, the computation of the full orientational correlation
function $g(\omega^2|r)$ and its subsequent integration is a
formidable task. In this specific case there are only four
independent degrees of freedom which can be fixed by providing,
e.g., the relative intermolecular distance $r$ and three
angles~\cite{soper}. Without loss of generality, we can hold the
center of mass of, say, particle $1$ fixed at the origin of the
laboratory coordinate system and let its molecular axis $\bf e_1$
lie along the $z$-axis. Then, the position and orientation of
particle $2$ can be unambiguously determined by specifying, for
instance, two polar coordinates of its center of mass, i.e., the
distance $r$ and the angle $\vartheta$ formed by the
intermolecular axis with the $z$-axis, and the Euler angles
$\theta$ and $\phi$ of the $\bf e_2$ axis (the ``twist'' angle,
$\chi$, is redundant for uniaxial molecules). In order to simplify
the whole scheme, we neglected the dependence of $g(\omega^2|r)$
on $\vartheta$ and $\phi$, thus focusing on the information
associated with the angle representing the relative orientation of
the two molecular axes, $\theta=\arccos{({\bf e_1}\cdot {\bf
e_2})}$. Indeed, this is the critical parameter that should be
monitored in order to assess the onset of either nematic or
smectic order in the fluid. The contracted orientational
distribution function $g(\theta|r)$ is normalized in such a way
that: \begin{equation} \frac{1}{2} \int_0^\pi g(\theta|r) \sin \theta {\rm d}
\theta = 1 \end{equation} We do not expect that such a simplification may
heavily bias the estimate of the weight associated with
orientational correlations in the {\it homogeneous} and {\it
isotropic} phase. Indeed, the quantity we are ultimately
interested in is just an integral measure that can help to assess
the relative balance between the positional and angular
contributions that intervene in the configurational entropy.

\section{Simulation}
We performed Monte Carlo (MC) simulations of hard spherocylinders
at constant pressure. This choice is rather common for systems
composed of non-spherical hard-core particles for which it may be
difficult to calculate the equation of state through the contact
value of the PDF, as is usually done in a constant-volume
simulation. Furthermore, density fluctuations are more naturally
accommodated in a constant-pressure simulation~\cite{smfr}. The
number of particles, that were enclosed in a cubic box with
periodic boundary conditions, typically ranged between $500$
($L/D=3.2,5$) and $800$ ($L/D=3$), with an intermediate value of
$672$ particles for $L/D=4$. A larger sample composed of $1500$
particles was used for more elongated particles with $L/D=20$.

A perfectly ordered, low-density lattice arrangement was the
starting configuration. Then, the system was allowed to
equilibrate, initially along a constant-volume MC path. A sequence
of states was then generated at constant pressure, starting from
an equilibrated isotropic configuration. We carried out rather
extended simulations, as compared with those currently reported in
the literature, especially in proximity of the transition
threshold. In fact, after an equilibration period of typically
$10^{5}$ to $10^{6}$ MC cycles, simulation data were obtained by
generating chains consisting of $5\times10^{5}$ to $5\times10^{6}$
MC cycles, depending on the pressure. A MC cycle consisted of an
attempt of changing, sequentially, both the position and the
orientation of each molecule, followed by an attempt of changing
the volume of the sample. During  such a procedure, each molecule
was first displaced from its original position, by moving at
random its center of mass within a cube. The molecule was then
re-oriented within a cone centered about the molecular axis. The
overall acceptance probability for the combined move was adjusted
to about 50\%. The isotropic change of the simulation box was
obtained by varying at random the box side $L$. Finally, the
maximum displacement allowed was set so as to achieve an average
acceptance probability for volume changes of $\sim 50$\%. For
hard-core molecules, the usual Metropolis acceptance criterion
reduces to a check of particle overlaps after each Monte Carlo
move. In this regard, a simple and efficient algorithm can be used
for spherocylinders~\cite{allen,vega}. During the cumulation runs,
we monitored the average density and built histograms for the
translational and orientational distribution functions,
respectively; equilibrium averages and standard deviations were
computed by dividing such runs into 5 to 10 independent blocks. In
the following, we shall refer to the volume packing fraction
$\eta=\rho V_{\rm hsc}$, where $V_{\rm
hsc}=(\pi/4)D^2L+(\pi/6)D^3$ is the volume occupied by one
spherocylinder, and to the reduced pressure $P^*=\beta PV_{\rm
hsc}$. We sampled the radial distribution function $g(r)$ at
intervals of $0.1D$, while the orientational distribution
$g(\theta|r)$ was calculated with a grid spacing of $\Delta \theta
= 5^\circ$.

We investigated the phase diagram of the model up to the threshold
of either the nematic or the smectic phase. The onset of
liquid-crystalline order was monitored through the nematic order
parameter: \begin{equation}\label{eq:p2} S=\frac{1}{N} \sum_{i=1}^N \
\frac{3\cos^2 \psi_i -1}{2}\;, \end{equation} where $\psi_i$ is the angle
formed by the molecular axis of the $i$-th molecule ${\bf e}_i$
(the sum running over all particles in the system). The director
${\bf n}$ defines the alignment direction attained, on average, by
the molecules. Since this direction is not known {\it a priori},
the order parameter is usually determined by diagonalizing the
traceless, symmetric, second-rank tensor $\bf Q$ that is defined
in terms of the molecular axes as: \begin{equation}\label{eq:q} {\bf Q}=
\frac{1}{N} \sum_{i=1}^N \ \frac{3{\bf e}_i{\bf e}_i-{\bf
I}}{2}\;, \end{equation} where ${\bf I}$ is the unit tensor. The nematic
order parameter is then obtained as the largest eigenvalue, and
the director as the corresponding eigenvector~\cite{allen}. The
value attained by $S$ is close to zero in the isotropic phase and
tends to one in a highly ordered phase. Across the ordering
transition, both the density and the order parameter show a finite
jump. However, such discontinuities, that are associated with a
first-order phase transition, are in certain cases so small that a
closer investigation of the correlation functions or a direct
observation of the arrangement of the molecules in the sample may
be necessary on the thermodynamic edge of both the nematic and
smectic phase.

As already anticipated, we used Widom's insertion
method~\cite{widom,smfr} to obtain independent numerical estimates
of the excess entropy at low densities, $s_{\rm ex}(\bar{\rho})$,
to be used in Eq.~(\ref{s:free_ex}). The Widom technique provides
a simple scheme for calculating the excess chemical potential,
$\mu_{\rm ex}$, of not too dense a liquid via the calculation of
the average Boltzmann factor associated with the (never accepted)
random insertion of an additional particle in an $N$-particle
system. The excess entropy can then be obtained through the
thermodynamic equation: \begin{equation}\label{eq:muex} s_{\rm ex} = - \beta
\mu_{\rm ex} + \frac{\beta P}{\rho} -1 \end{equation} Depending on the
density and the elongation of the spherocylinders, 50 to 200 trial
insertions per MC step were typically needed in order to obtain a
reliable estimate of $\mu_{\rm ex}$. The use of this technique at
small but finite densities avoids the otherwise necessary
extrapolation to the ideal gas limit, as a reference state for the
integration contemplated in Eq.~(\ref{s:free_ex}). This latter
procedure was adopted in~\cite{csg}, where some spurious
fluctuations of the orientational distribution function as well as
some difficulties associated with the simulation in the highly
dilute regime were registered.

As sketched in Fig.~\ref{f1}, the values obtained for $s_{\rm ex}$
using Widom's insertion method are quite stable in all the cases
we have examined: the thermodynamic potential relaxes almost
immediately, with small fluctuations about the average value. No
significant drifts were observed during simulations as long as
$10^5$ MC steps. The simulation parameters and the results for the
relevant properties of the fluid at low densities were collected
for various elongations in Table~\ref{t1}.

\section{Results and discussion}
\subsection{Phase behavior for $3 \leq L/D \leq 5$}
In this subsection we shall focus on the phase behavior of the
model in a range of parameters where, for increasing values of the
aspect ratio, the disordered fluid undergoes a transition to
either a solid, a smectic or a nematic phase. We investigated the
model for shape anisotropies $L/D=3, 3.2, 4,$ and $5$. For
$L/D=3$, the fluid freezes into a crystalline phase. An I-SmA
phase transition is observed over a range of larger aspect ratios
bounded by an isotropic-smectic-solid (I-SmA-S) triple point
($L/D=3.1$) and by an I-N-SmA triple point ($L/D=3.7$),
respectively. For even larger anisotropies, the disordered fluid
forms, under compression, a nematic phase~\cite{bofr}.

The reliability of the present results for the EoS and for the
nematic order parameter can be appraised through Fig.~\ref{f2}
where the above quantities were plotted as a function of the
packing fraction for $L/D=3.2$ and $5$. The comparison with the
simulation data reported by McGrother and co-workers~\cite{jawil}
shows that a system with $500$ particles is fairly adequate for
evaluating the basic thermodynamic properties of the fluid. The
agreement turned out to be equally good along the disordered
branch also for $L/D=3$ and $4$, up to the transition point.

We now turn to a discussion of the properties of the model for
$L/D=5$, a case where the isotropic fluid undergoes a rather weak
first-order transition to a thermodynamically stable nematic phase
at about $45\%$ of the close-packing density, a value
corresponding to an absolute packing fraction of
$0.4$~\cite{jawil,bofr}. No specific feature indicating the
transition edge is apparent either in the equation of state or in
the order parameter (see Fig.~\ref{f2}). A rather modest degree of
positional order shows up in the spatial profile of the radial
distribution function that was plotted in Fig.~\ref{f3} for
increasing pressures, all the way up to the transition threshold.
Actually, for moderately small packing fractions, the local
density around one generic spherocylinder is significantly
reduced, with respect to the homogeneous value taken at large
distances, for interparticle separations lower than
$\sim3\frac{1}{2}$ molecular diameters, a distance which
corresponds to the minimum contact separation between two
orthogonal spherocylinders.

Retrospectively, this local ``depletion'' effect is more easily
understood if one looks at the orientational distribution function
(see Fig.~\ref{f4}): the almost flat profile of $g(\theta|r)$ for
$r>3.5D$, when plotted as a function of $\theta$, implies that
orthogonal geometries in the relative spatial arrangement of two
spherocylinders are equally probable as parallel configurations
are. The steric hindrance produced, on average, by such orthogonal
(or almost orthogonal) particles sensitively reduces the
probability of finding other molecules at shorter distances from
the central one. Upon further compressing the fluid, the value of
$g(\theta|r)$ for aligned-particles geometries ($\theta=0,\pi$)
steadily increases while different geometries become, at the same
time, rarer and rarer. This is apparent from the progressive and
systematic reduction of $g(\theta|r)$, for short and intermediate
molecular separations, over a wide range of angles centered about
$\pi/2$. Correspondingly, the radial distribution function becomes
more and more structured at short distances as is distinctly shown
in Fig.~\ref{f3} by the building up of a first coordination shell
formed by particles lying almost at close contact with the central
one.

The excess entropy of the fluid was resolved in Fig.~\ref{f5} in
terms of the contributions associated with pair and
more-than-two-particle correlations, respectively (see
Eq.~(\ref{s:sex})). The behavior of the RMPE is similar to that
exhibited by the same quantity in a variety of other model
systems, composed of spherical particles, across an ordering
transition such as the freezing of a fluid~\cite{spg1,spg2}
 or the phase separation
of a binary mixture~\cite{sg}. In particular, this quantity
vanishes for a value of the packing fraction ($\eta_0$) which
falls precisely on top of the currently estimated I-N phase
transition point (see Table~\ref{t2} where we summarized the
predictions of the entropy-based criterion for the presently
investigated values of the aspect ratio). We note that the current
RMPE estimate sensitively improves on that formerly given
in~\cite{csg} where the values reported for the pair entropy
turned out to be affected by a systematic computational error.
However, no significant morphological change resulted, {\it a
posteriori}, in both the pair entropy and in the RMPE. We also
recall that the Authors had shown in~\cite{csg} that the pair
entropy changes with the size of the sample, for a given density,
in a tiny but systematic way, the zero-RMPE density shifting
towards moderately larger values. The RMPE inverts its trend (from
decreasing to increasing) in the density region ($\eta \sim 0.36$)
corresponding to a more extended structuring of the fluid at short
distances (see Fig.~\ref{f3}). To be specific, we refer to the
growing of a local nucleus of (mostly) aligned particles that is
signalled by the emergence of a secondary maximum in the radial
distribution function at a relative distance twice as large as
that corresponding to the first coordination shell.

Figure~\ref{f5} also shows the pair entropy of the fluid resolved,
according to Eq.~(\ref{s:s2}), into the sum of a (spatially
independent) excluded-volume term ($-B_2\rho$) plus two more
contributions arising from translational and orientational
correlations, respectively. As expected, it is $s_2^{\rm or}$ that
drives the rather sharp increase of $\Delta s (\eta)$ just beyond
the minimum. {\it Vice versa}, the contribution associated with
positional correlations, $s_2^{\rm tr}$, is very small all over
the fluid range. The function which, upon integration (see
Eq.~(\ref{s:s2or})), yields the quantity $s_2^{\rm or}$ is shown,
for increasing densities, in the right part of Fig.~\ref{f3}. The
value of the corresponding integral is negative and keeps close to
zero at low and intermediate densities (see Fig.~\ref{f5}). For
packing fractions larger than $\sim 0.3$, $s_2^{\rm or}$ starts
decreasing more and more rapidly and eventually overcomes (in
absolute value) the excluded-volume term. As seen from
Fig.~\ref{f5}, the fingerprint of the ordering of the fluid is
manifest in the persistence of angular correlations up to
intermediate and long distances.

Considerations similar to those developed above for $L/D=5$ apply
also to the $L/D=3.2$ case, even if it is a smectic phase that is
now formed by the isotropic fluid. However, the basic underlying
mechanism which drives the first-order I-SmA transition is
analogous to that exploited in the nematic case. Obviously, the
denser isotropic fluid looks more structured than for $L/D=5$ (see
Fig.~\ref{f6}). Furthermore, at the transition point angular
correlations show a very slow spatial decay. We also note that in
this case the lower molecular asymmetry modifies the scale of
interparticle distances in the average density profile. Thus, the
effect associated with the structuring of the second coordination
shell, which we already commented upon for $L/D=5$, is less
resolved since the corresponding distance falls very close to the
position of the lone weak maximum, located at a distance $r/D=1+
\frac{1}{2}(L/D)=2.6$, that is present in the radial distribution
function at low densities. A shoulder, rather than a peak, emerges
at $r/D\simeq 2.4$, again in a density range corresponding to the
position of the minimum ($\eta \simeq 0.47$) attained by the RMPE
(see Fig.~\ref{f7}).

In general, the ordering thresholds detected through the vanishing
of the RMPE correlate very well with the corresponding transition
point (see Table~\ref{t2}), whatever the nature of the
higher-density phase coexisting with the isotropic fluid. The
evidence discussed so far justifies the claim that the RMPE of
spherocylinders monitors the leading microscopic process which
drives the phase transition in all such cases, i.e., the aligning
of spherocylinders along a common direction. As such, the
vanishing of this quantity presumably portends the density
threshold beyond which the nematic ordering may possibly emerge in
the system, regardless of other concurrent structural phenomena
like those which may ultimately lead to the formation of a more
ordered macroscopic state, be it smectic or crystalline, instead
of a pure nematic phase. An indication consistent with this thesis
that is, perhaps, not accidental (yet to be considered with due
caution because of the numerical uncertainty arising from the
computational error which affects, at least, the third significant
figure) is offered by the apparently greater accuracy of the
predictions given by the RMPE-based criterion for $L/D=4$ and $5$
(see Table~\ref{t2}), viz., in those cases where the stable
coexisting phase is a truly nematic fluid. In the other two cases
investigated, i.e., when either a solid or a smectic phase is
formed, the criterion just overshoots the target, locating the
internal threshold for the ordering of the fluid at a slightly
higher density.

\subsection{The Onsager regime}
As the aspect ratio increases, the I-N transition point moves to
lower and lower densities. In the Onsager limit of infinitely long
rods ($L/D\to\infty$), a fluid of hard spherocylinders undergoes a
first-order phase transition to a nematic phase for
$B_2\rho=3.29$~\cite{ons,lek,wj}. However, the phase behavior
characteristic of the Onsager regime qualitatively sets in for
shape anisotropies even less than $L/D=20$, even if for such
values of the aspect ratio the basic assumption made by Onsager
that all virial coefficients of order higher than two can be
neglected is not yet satisfied~\cite{bofr}. In fact, the
contribution of the third virial term to the free energy,
evaluated for $L/D=20$ at the transition density, turns out to be
still comparable to the contribution of the second virial
term~\cite{straley,frenkel2}.

We carried out simulations for $L/D = 20$. In passing, we recall
that this value mimics the typical size of the tobacco mosaic
virus, whose particles have a practically rigid rod shape of
$15-18$ nm diameter and $300$ nm length~\cite{uist}. For such
large values of the aspect ratio, the simulation of a fluid of
hard spherocylinders dictates more severe constraints on the size
of the simulation box which should be large enough to accommodate
at least two rod lengths, so as to avoid multiple overlaps between
a molecule and the periodic images of another
particle~\cite{bofr}. The number of particles chosen ($N=1500$) is
sufficient to ensure that the condition mentioned above is
fulfilled in the (relatively low) density range where the
disordered phase is thermodynamically stable. The reduced pressure
and the nematic order parameter are shown in Fig.~\ref{f8}. The
isotropic liquid phase is mechanically stable up to a reduced
pressure $P^*=0.92$: in this regime the system shows, for
simulations as long as $2.5\times 10^6$ MC cycles, no significant
fluctuations about the average values of the packing fraction
($\langle\eta\rangle=0.145$) and of the nematic order parameter
($\langle S \rangle =0.058$), respectively. Upon further
compressing the system up to $P^*=0.93$, we observed, after about
$1.5\times 10^6$ MC cycles, a spontaneous and rapid transformation
into a nematic phase which, at equilibrium, settled down at a
packing fraction $\langle \eta \rangle=0.182$ and was
characterized by an order parameter $\langle S \rangle=0.862$. The
resulting ``boundaries'' are slightly shifted toward higher
densities if compared with the direct estimates reported
in~\cite{bofr}, i.e., $\eta_{\rm iso}=0.139$ or $0.137$, and
$\eta_{\rm nem}=0.171$ or 0.172, according to whether the
Gibbs-Duhem integration or a Gibbs-ensemble technique was used.
Exploratory runs performed by decompressing the system from the
ordered phase distinctly show an hysteresis, the fluid remaining
almost nematic for pressures as low as $P^* \sim 0.75$. We
calculated the pressure also using the ``decoupling
approximation'' between the translational and orientational
degrees of freedom originally introduced by
Parsons~\cite{parsons}. In this approximation a system of rods is
mapped onto a system of spheres interacting with an effective
orientation-dependent potential. As already observed by McGrother
and co-workers for the shorter elongation regime $3.2 \leq L/D
\leq 5$ that was analyzed in~\cite{jawil}, the resulting EoS turns
out to be fairly accurate in the isotropic phase up to the
transition point, and definitely better than the prediction based
on the scaled-particle-theory approach exploited by
Boublik~\cite{boublik}. However, it should be noted that the
latter approach is based on a one-parameter approximation for the
third and fourth virial coefficients which cannot be expected to
yield accurate results for large elongations~\cite{czech}. A
better agreement with the simulation data is obtained with a
closely related description that was proposed by
Nezbeda~\cite{czech,nez}: this approximation is based on an
empirical fit of the virial coefficients, especially tailored on
the available results for hard spherocylinders. All the above
approximations for the EoS were reported, for a comparison, in
Fig.~\ref{f8}: note that the simulation data are closely bracketed
by the Parsons and Nezbeda approximations, respectively.

Radial correlations are shown in Fig.~\ref{f9} together with the
function whose space integral yields $s_{2}^{\rm or}$. All over
the liquid density range, the radial distribution function is
practically structureless apart from the correlation hole at short
distances (see Fig.~\ref{f9}). As a result, the translational
contribution to the pair entropy keeps close to zero all over the
explored density range (see Fig.~\ref{f10}). The same is also true
for the orientational contribution for packing fractions less than
about $0.15$. It thus follows that the pair entropy of the
isotropic fluid is substantially equal to the excluded-volume term
$-B_2\rho$, as is actually assumed in the original Onsager theory.
At variance with the positional term, the angular contribution is
responsible, together with the excluded-volume term, of the
vanishing of the RMPE. Just beyond the transition point, $s_2^{\rm
or}$ blows up and definitely overcomes the excluded-volume term
(see Fig.~\ref{f10}). The resulting RMPE shows the typical
behavior already observed for smaller aspect ratios. Also in this
case, the zero-RMPE threshold is in very good agreement with the
I-N transition point estimates given in ~\cite{bofr} (see
Table~\ref{t2}). We also reported in Fig.~\ref{f10} the Parsons
and Nezbeda analytical estimates for the excess entropy. As
already pointed out for the EoS, both approximations turn out to
predict fairly well, almost with the same degree of accuracy, also
the excess entropy of the fluid.

As seen from Figs.~\ref{f5},~\ref{f7} and~\ref{f10}, the overall
importance of the cumulative contribution of higher-order density
correlations to the excess entropy decreases with increasing shape
anisotropies. This is clearly a consequence of the progressive
confinement of the isotropic phase to lower and lower densities.
In this regard, we just observe that the value attained by the
RMPE at its minimum drops from $\sim 45\%$ of the excess entropy
for $L/D=3.2$ to $\sim 18\%$ for $L/D=20$. This behavior is
consistent with the numerical findings on the virial expansion
cited above and with the asymptotic trend predicted by Onsager as
$L/D\to\infty$. Note, however, that at low densities the ratio
$\Delta s/s_{\rm ex}$ moderately increases with $L/D$ for a given
packing fraction. Actually, longer and thinner particles give rise
to more extended multiparticle correlations which trigger the
ordering of the fluid at lower and lower densities. Beyond the
zero of the RMPE, where the nematic phase is thermodynamically
stable, the present data monitor {\it de facto} a system in a
metastable, macroscopically isotropic condition whose proper
description lies beyond the standard Onsager theory.

\section{Concluding remarks}
In this paper we have analyzed the contributions to the ``pair
entropy'' $s_2$ of hard spherocylinders of length $L$ and diameter
$D$ that are associated with excluded-volume, translational and
angular correlations, with particular emphasis on the
thermodynamic behavior exhibited by the above quantities on
approaching the phase transition from the low-density fully
disordered phase to a partially ordered phase at higher densities.
An elucidating synthesis of the results obtained for $s_2$ through
an intensive Monte Carlo sampling of the model is offered by
Fig.~\ref{f11} where we plotted the three contributions cited
above, each of them being calculated in the state corresponding to
the vanishing of the residual multi-particle entropy (RMPE) for an
assigned value of $L/D$. The RMPE is a quantity that gauges the
cumulative weight of spatial correlations involving {\it more than
two particles} in the configurational entropy of the fluid.
Figure~\ref{f11} neatly indicates the relative importance of the
three contributions in the overall entropic balance: for very
small asymmetries, the translational term, $s_2^{\rm tr}$,
prevails (in absolute value) while being comparable with the
excluded-volume contribution, $-B_2\rho$ (note that the data for
$L/D=0$ do obviously refer to the hard-sphere case~\cite{gg}). For
increasing values of the aspect ratio the angular term, $s_2^{\rm
or}$, acquires more and more importance: in fact, an interpolation
of the available data shows that the three contributions reach the
same magnitude for $L/D\simeq0.4$. Actually, it is in this range
of values (more precisely, for $L/D\leq0.35\pm0.05$~\cite{bofr})
that the ``rotator phase'' (i.e., the plastic crystal) disappears
in the phase diagram of hard spherocylinders, and is replaced at
high densities by an orientationally ordered, crystal phase. As
seen from Fig.~\ref{f11}, in the range $0.4\lesssim{L/D}\lesssim5$
the ordering of the fluid is mainly controlled by angular
correlations. In fact, the translational contribution rapidly
tends to smaller and smaller values and practically vanishes for
$L/D>5$. Also $s_2^{\rm or}$ asymptotically tends to zero in this
range of increasing asymmetries but, as seen from Fig.~\ref{f11},
the decay of this quantity is much slower than that shown by its
translational counterpart. Correspondingly, the excluded-volume
term becomes the dominant contribution to the configurational
entropy of the fluid. Note that its value for $L/D=20$ ($-3.45$)
is already very close to the Onsager limit ($-3.29$).

As far as the quality of the ``predictions'' offered by the
zero-RMPE criterion are concerned, the agreement with the properly
defined thermodynamic thresholds is impressive for all the
asymmetries that have been investigated. Indeed, the vanishing of
the RMPE monitors, in a very sensitive and reliable way, the
ordering of the homogeneous and isotropic fluid into a more
ordered phase be it nematic, smectic or solid. However, we remark
that such a close quantitative correspondence should not be
necessarily and systematically expected {\it a priori}. In fact,
the zero of $\Delta s$ needs not coincide with the thermodynamic
edge of the disordered phase that is properly located through the
comparison of the free energies of the two coexisting phases.
Actually, the RMPE conveys a type of information that is
intermediate between the fully macroscopic level and the
underlying microscopic description of the system. The
interpretation of such a simply accessible information (and, more
specifically, of its change of sign) as an intrinsic signature of
the incipient ordering of the fluid appears to be quite firmly
established on the basis of an ample and coherent evidence that
has, by now, emerged in a variety of continuum and lattice model
fluids~\cite{spg1,sg,spg2}. The present analysis of the ``fine
structure'' of the configurational entropy of hard spherocylinders
further corroborates the sensitivity of the RMPE to local ordering
phenomena which may even turn out into a rather weak first-order
phase transition. We believe that this one-phase entropic
criterion can be conveniently generalized to account for more than
one phase transition, leading the system to increasing levels of
self-organization. An attempt in this direction has been recently
done by Cuetos and coworkers whose results seem to indicate the
existence of a jump in the translational part of the pair entropy
as well as in its rate of change across the nematic-smectic
transition~\cite{cuetos}. However, this evidence was not
considered conclusive because of the limited size of the
simulation box. We are also currently investigating this point
with promising preliminary results.

\section*{Acknowledgments}

We would like to thank Professor Bruno Martinez-Haya for useful
discussions and for pointing a numerical inconsistency out in a
previously published article. D.~C. gratefully acknowledges the
hospitality and advice of Professor M.~P.~Allen during a visit to
the University of Bristol.

\clearpage


\section*{Tables}

\begin{table}[h!]
\caption{Thermodynamic properties of the fluid in the low-density
states $\bar\rho$ used for the integration of the equation of state in
Eq.~(\ref{s:free_ex}) of the text. Column VIII refers to the number
of trial insertions per MC cycle; the percentage ratio of accepted
trial insertions is reported in column IX. Standard
deviations in the last digit(s) are given in parenthesis.} \label{t1}
\bigskip
\begin{tabular}{lllllllcl} \\
$L/D$ & $P^*$ & $\langle\eta\rangle$
& $\langle\beta\mu_{\rm ex}\rangle$ & $\langle s_{\rm ex}\rangle$
& $\langle s_2\rangle$ & MC cycles & Trials & Ratio (\%)  \\  \\
\hline \\
3   & 0.30 & 0.131(1) & 2.33(1)& -1.039(13)  & -0.959(2)
    & $5\times10^5$ &$100$& 9.7\\
3.2 & 0.30 & 0.130(1) & 2.37(1)& -1.054(17)  & -0.971(3)
    & $9\times10^5$ &$50\div200$& 9.4\\
4   & 0.30 & 0.125(1) & 2.55(1)& -1.149(6)  & -0.996(1)
    & $5\times10^5$ &$80\div200$& 7.8\\
5   & 0.181 & 0.089(1) & 1.90(1)& -0.878(6) &   -0.834(1)
    & $6\times10^5$ &20& 14.9\\
    & 0.379 & 0.136(1) & 3.25(1)& -1.457(24) &  -1.237(4)
    & $5\times10^5$ &200& 3.8\\
20  & 0.10 & 0.044(1) & 2.39(1) & -1.137(13) & -1.066(1)
    & $5\times10^5$ & 50  &  9.2 \\
    & 0.20 & 0.066(1) & 3.79(1) & -1.785(26) & -1.588(5)
    & $10\times10^5$ & 200  &  2.2 \\  \\
\end{tabular}
\end{table}

\begin{table}
\caption{The packing fractions corresponding to the vanishing of
the residual multiparticle entropy are compared, for different
elongations, with the thermodynamic thresholds of the phase
transitions undergone by the isotropic fluid, according to the
available simulation data. Note that the value ascribed to
Ref.~\protect\cite{bofr} for $L/D=4$ was estimated through a
linear interpolation of the data for the I-N transition densities
reported in Ref.~\protect\cite{bofr} for $L/D=3.8$ and $L/D=4.2$,
respectively. A slightly lower estimate for the I-N transition
density at $L/D=4$, $\eta=0.459$, is reported in
Ref.~\protect\cite{alle}.} \label{t2}
\bigskip
\begin{tabular}{lcccc}
\\ & & &\multicolumn{2}{c} {Phase transition point} \\
\cline{4-5}\\
$L/D$ & Phase transition & {$\eta_0$} &
{Ref.~\protect\cite{jawil}} & {Ref.~\protect\cite{bofr}} \\ \\
\hline \\
3 & I-S   & 0.521 & 0.490 & 0.512 \\
3.2 & I-SmA & 0.510 & 0.513 & 0.503 \\
4 & I-N   & 0.463 & 0.472 & 0.466 \\
5 & I-N   & 0.399 & 0.407 & 0.398 \\
20  & I-N   & 0.148 &  ---  & 0.139\\ \\
\end{tabular}
\end{table}

\clearpage


\section*{Figures}

\begin{figure}[h]
\caption{Typical evolution of $s_{\rm ex}(\bar\rho)$ (see
Table~\ref{t1}) during a MC run for different elongations. The
data were obtained using the Widom test method. Circles: $L/D=3.2$
at $P^*=0.30$; diamonds: $L/D=5$ at $P^*=0.379$; squares: $L/D=20$
at $P^*=0.20$. } \label{f1}
\end{figure}

\begin{figure}[h]
\caption{Equation of state (top) and nematic order parameter
(bottom) for $L/D = 3.2$ (left panels) and 5 (right panels). Solid
circles: this work; triangles: MC simulations by McGrother {\it et
al.}~\protect\cite{jawil}. The error bars are systematically
smaller than the size of the markers.} \label{f2}
\end{figure}

\begin{figure}[h]
\caption{Radial distribution function (left) and integrand (right)
for $s_2^{\rm or}$ in Eq.~(\ref{s:s2or}) at several densities for
$L/D=5$. Open triangles: $P^*=0.18$, $\langle\eta\rangle=0.089$;
solid circles: $P^*=1.10$, $\langle\eta\rangle=0.223$; squares:
$P^*=2.53$, $\langle\eta\rangle=0.310$; solid triangles:
$P^*=4.20$, $\langle\eta\rangle=0.372$; diamonds: $P^*=4.95$,
$\langle\eta\rangle=0.397$. }\label{f3}
\end{figure}

\begin{figure}
\caption{Orientational distribution functions $g(\theta|r)$ at
$P^*=1.10$ (left) and 4.95 (right), plotted, from top to bottom,
for several distances ($r/D=$ 1.25, 1.95, 2.55, 3.65, 6.05).}
\label{f4}
\end{figure}

\begin{figure}[h]
\caption{Left panel: residual multiparticle entropy (circles)
resolved into the excess (squares) and pair (solid diamonds)
contributions for $L/D=5$; the values of the excess entropy at
lower densities were also computed using the Widom test method
(pluses). Right panel: pair entropy (solid diamonds) resolved into
the translational (triangles), orientational (squares), and
excluded-volume (circles) contributions, according to
Eq.~(\ref{s:s2}). Lines are smooth interpolations of the
simulation data.} \label{f5}
\end{figure}

\vspace{2cm}

\begin{figure}[t]
\caption{Radial distribution function (left) and integrand (right)
for $s_2^{\rm or}$ in Eq.~(\ref{s:s2or}) at several densities for
$L/D=3.2$. Open triangles: $P^*=0.5$, $\langle\eta\rangle=0.170$;
     solid circles:       $P^*=2.0$, $\langle\eta\rangle=0.305$;
     squares:       $P^*=5.0$, $\langle\eta\rangle=0.411$;
     solid triangles:   $P^*=8.0$, $\langle\eta\rangle=0.474$;
     diamonds:      $P^*=9.8$, $\langle\eta\rangle=0.507$;
pluses (right panel): $P^*=9.9$, $\langle\eta\rangle=0.511$.
}\label{f6}
\end{figure}

\begin{figure}[t]
\caption{Left panel: residual multiparticle entropy (circles)
resolved into the excess (squares) and pair (solid diamonds)
contributions for $L/D=3.2$; the excess entropy at the lowest
packing fraction was computed using the Widom test method. Right
panel: pair entropy (solid diamonds) resolved into the
translational (triangles), orientational (squares), and
excluded-volume (circles) contributions, according to
Eq.~(\ref{s:s2}). Lines are smooth interpolations of the
simulation data.} \label{f7}
\end{figure}

\begin{figure}[t]
\caption{Left panel: equation of state for $L/D = 20$ (solid
circles); the Parsons (dotted line,~\protect\cite{parsons}),
Boublik (solid line,~\protect\cite{boublik}), and Nezbeda (dashed
line,~\protect\cite{nez}) approximations for the equation of state
are also shown. Right panel: nematic order parameter plotted as a
function of $\langle \eta \rangle$. Triangles refer in both panels
to states obtained upon decompressing the nematic fluid. The error
bars are systematically smaller than the size of the
markers.}\label{f8}
\end{figure}

\begin{figure}[t]
\caption{Radial distribution function (left) and integrand (right)
for $s_2^{\rm or}$ in Eq.~(\ref{s:s2or}) at several densities for
$L/D=20$. Circles:   $P^*=0.10$, $\langle\eta\rangle=0.044$;
squares:   $P^*=0.50$, $\langle\eta\rangle=0.107$; triangles:
$P^*=0.85$, $\langle\eta\rangle=0.140$; solid diamonds:
$P^*=0.93$, $\langle\eta\rangle=0.182$; pluses (right panel):
$P^*=0.70$, $\langle\eta\rangle=0.127$. }\label{f9}
\end{figure}

\begin{figure}[t]
\caption{Left panel: residual multiparticle entropy (circles)
resolved into the excess (squares) and pair (solid diamonds)
contributions for $L/D=20$; the excess entropy at lower densities
was also computed using the Widom test method (pluses). The
Parsons (dotted line,~\protect\cite{parsons}) and Nezbeda (dashed
line,~\protect\cite{nez}) approximations for the excess entropy
are also shown. Right panel: pair entropy (solid diamonds)
resolved into the translational (triangles), orientational
(squares), and excluded-volume (circles) contributions, according
to Eq.~(\ref{s:s2}). Solid lines are smooth interpolations of the
simulation data. }\label{f10}
\end{figure}

\begin{figure}[t]
\caption{ Translational (circles), orientational (squares), and
excluded-volume (solid diamonds) contributions to the pair entropy
of the fluid, calculated in the states of vanishing residual
multiparticle entropy, for different shape anisotropies. Data for
$L=0$ (hard spheres) were taken from Ref.~\protect\cite{gg}; data
for $L=1$ were obtained from the equation of state reported in
Ref.~\protect\cite{verfre}.} \label{f11}
\end{figure}

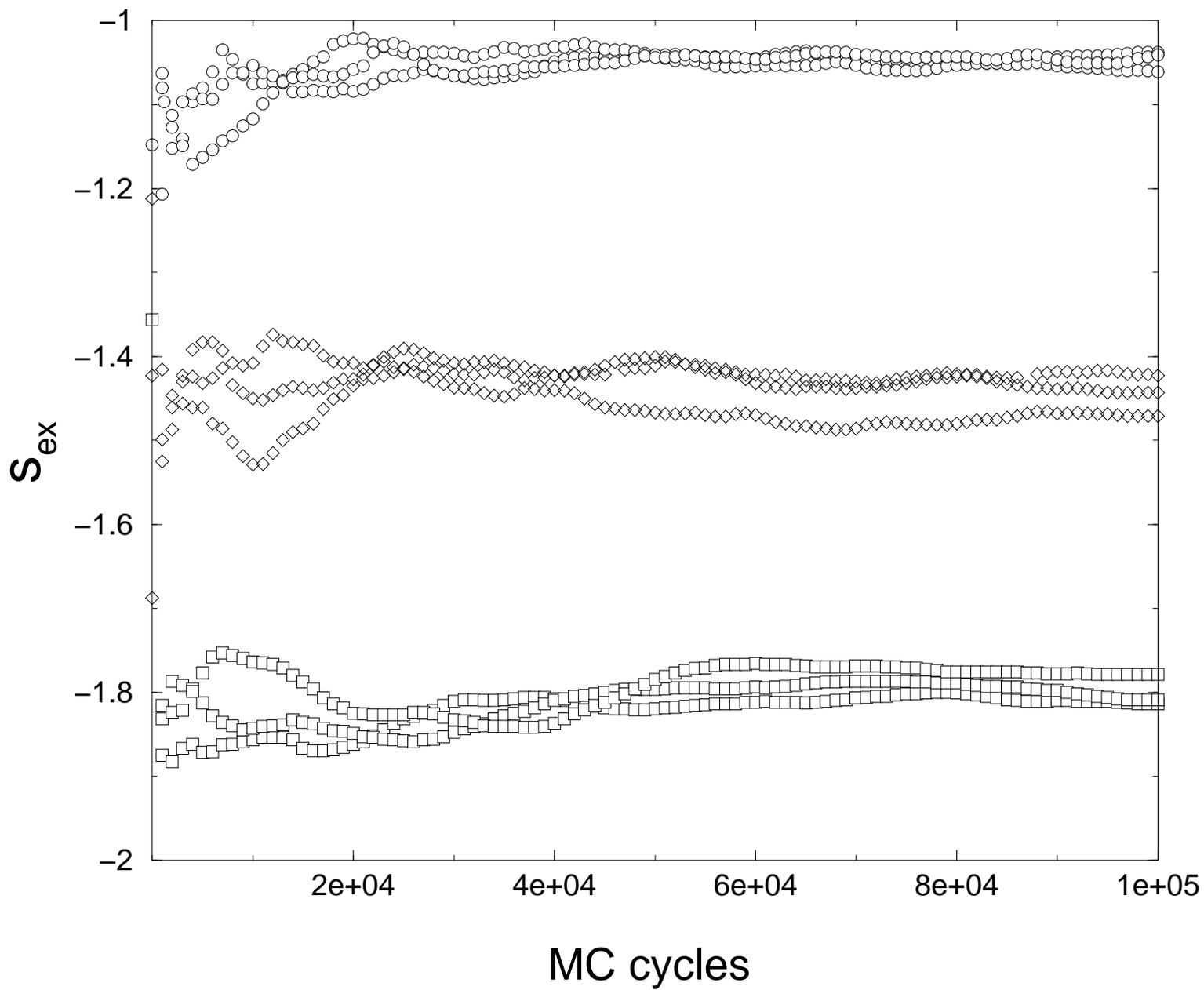
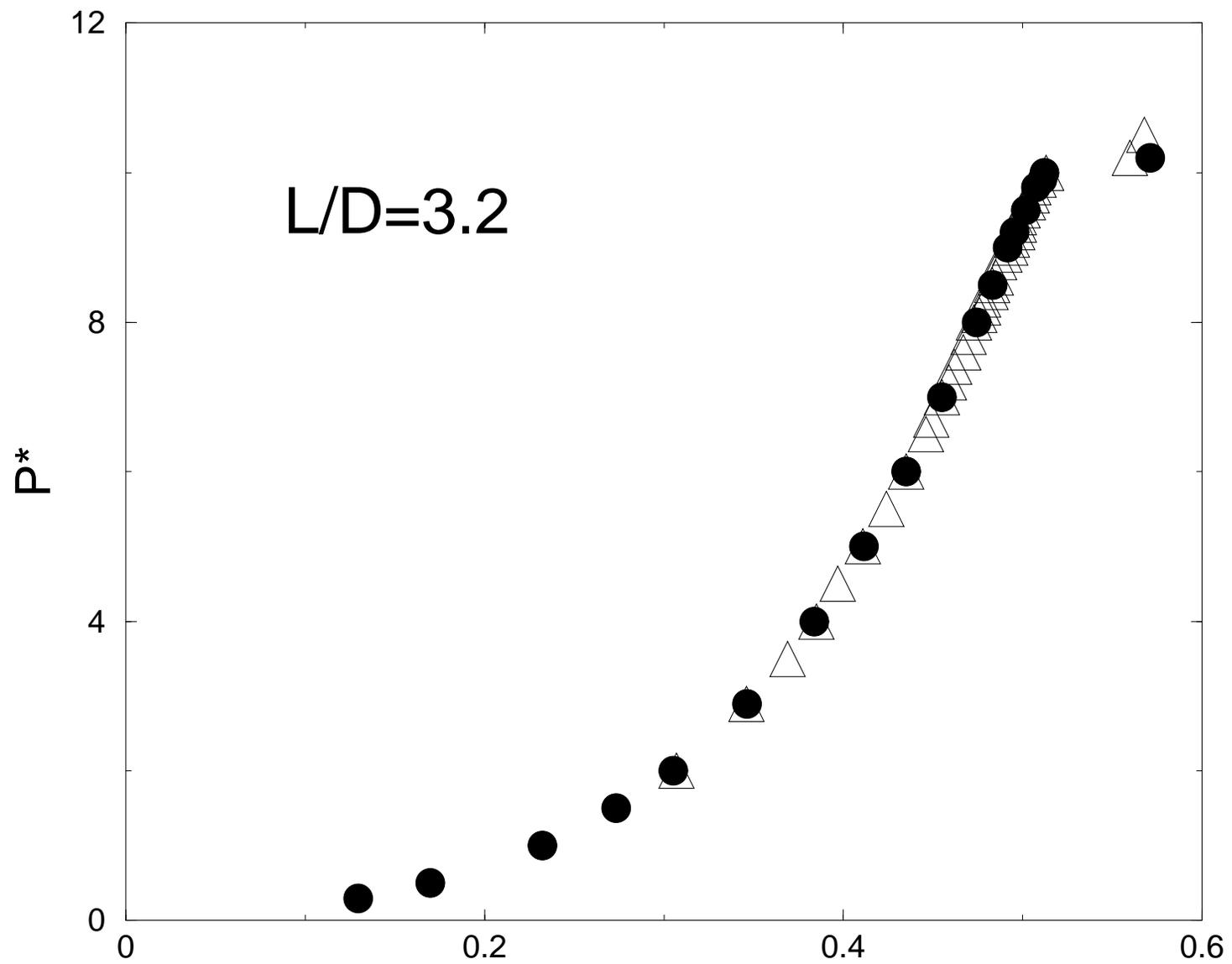
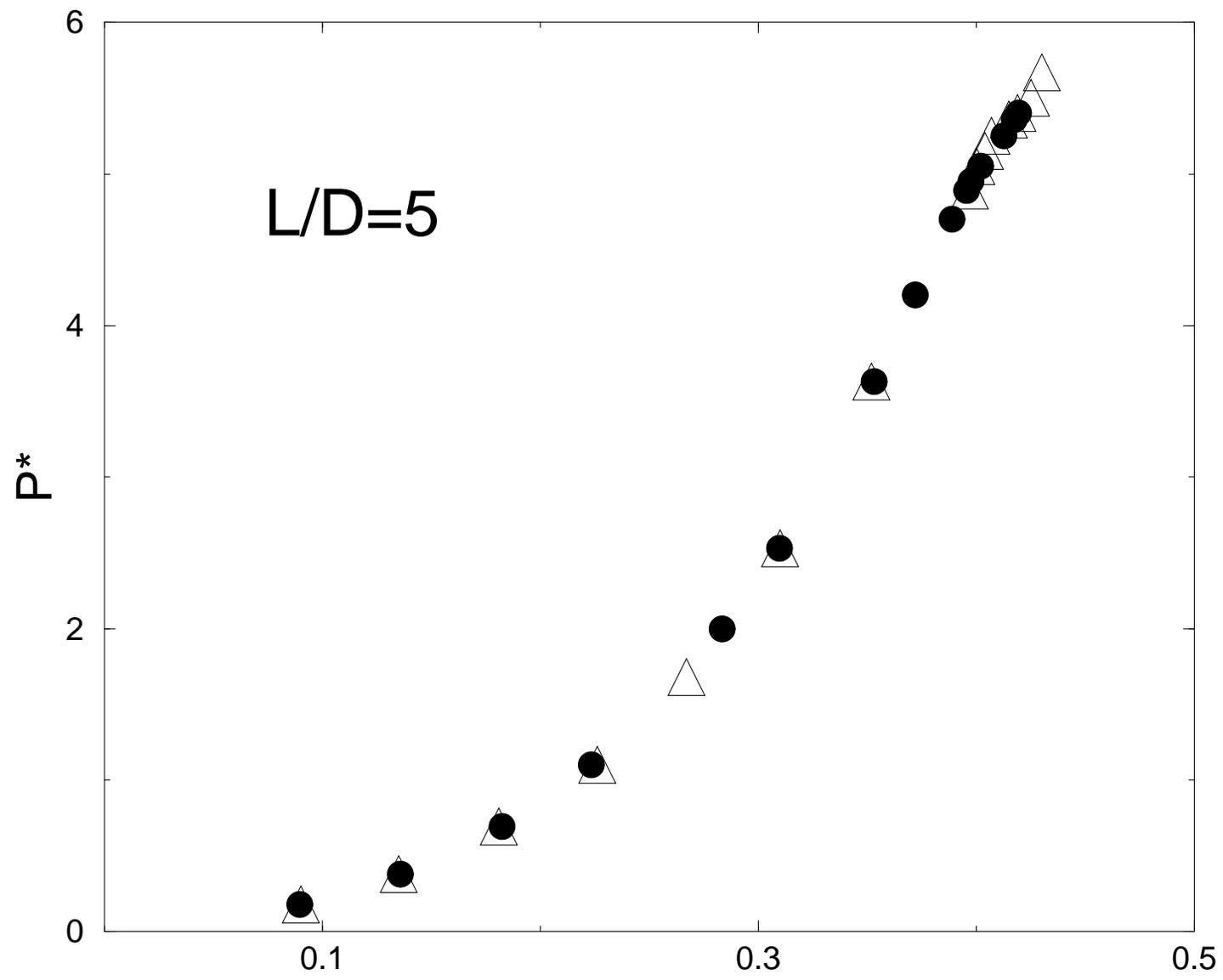
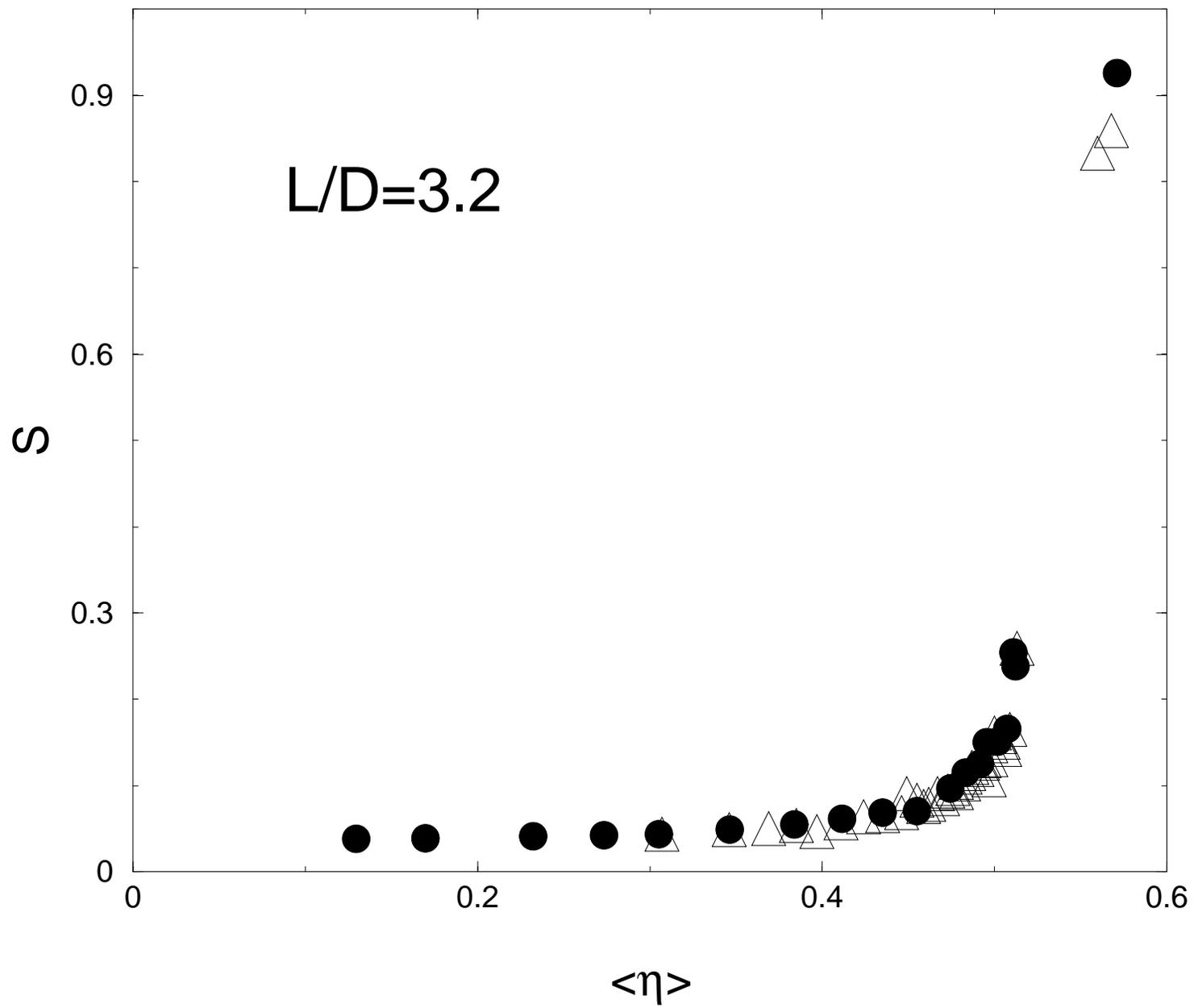
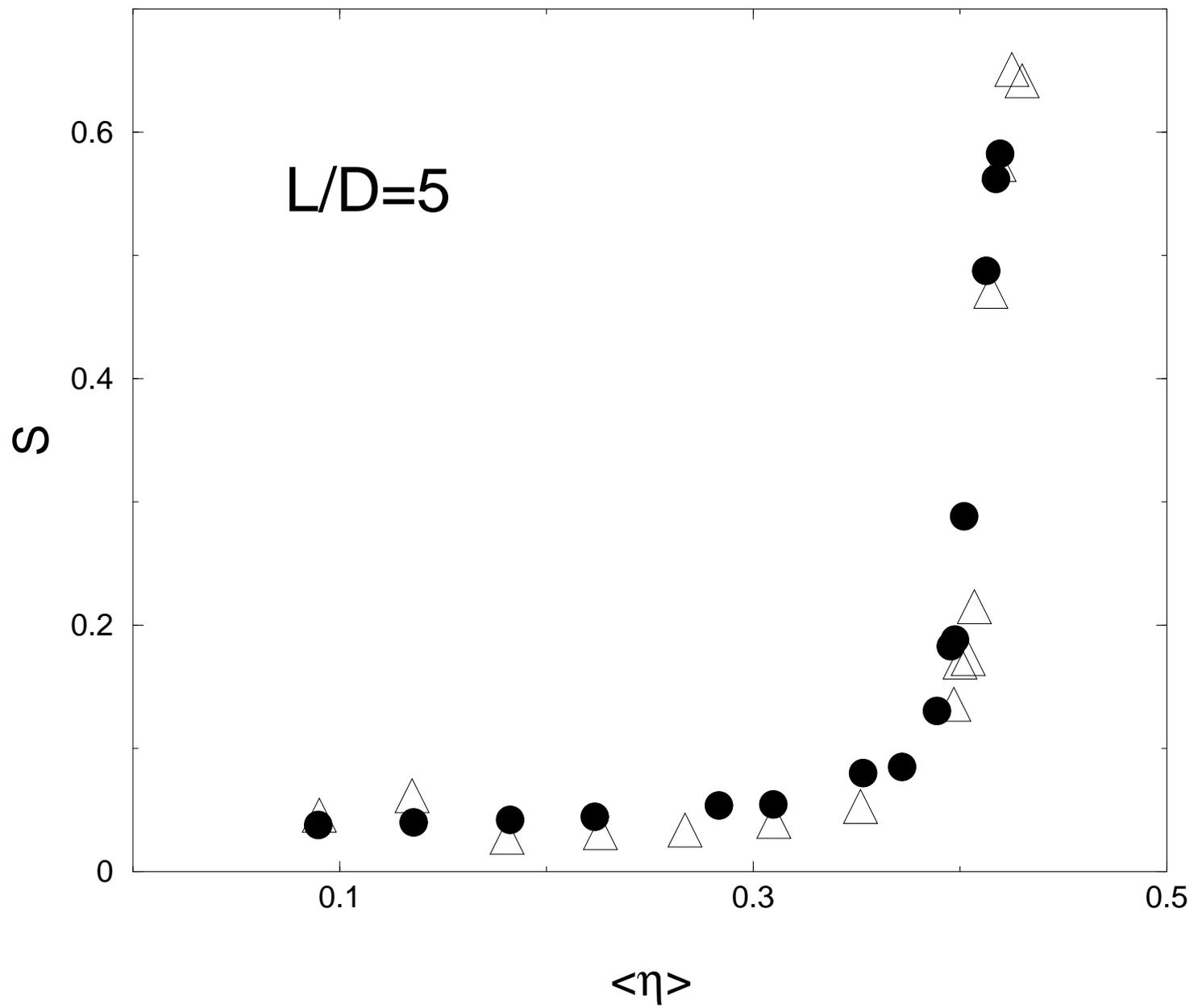
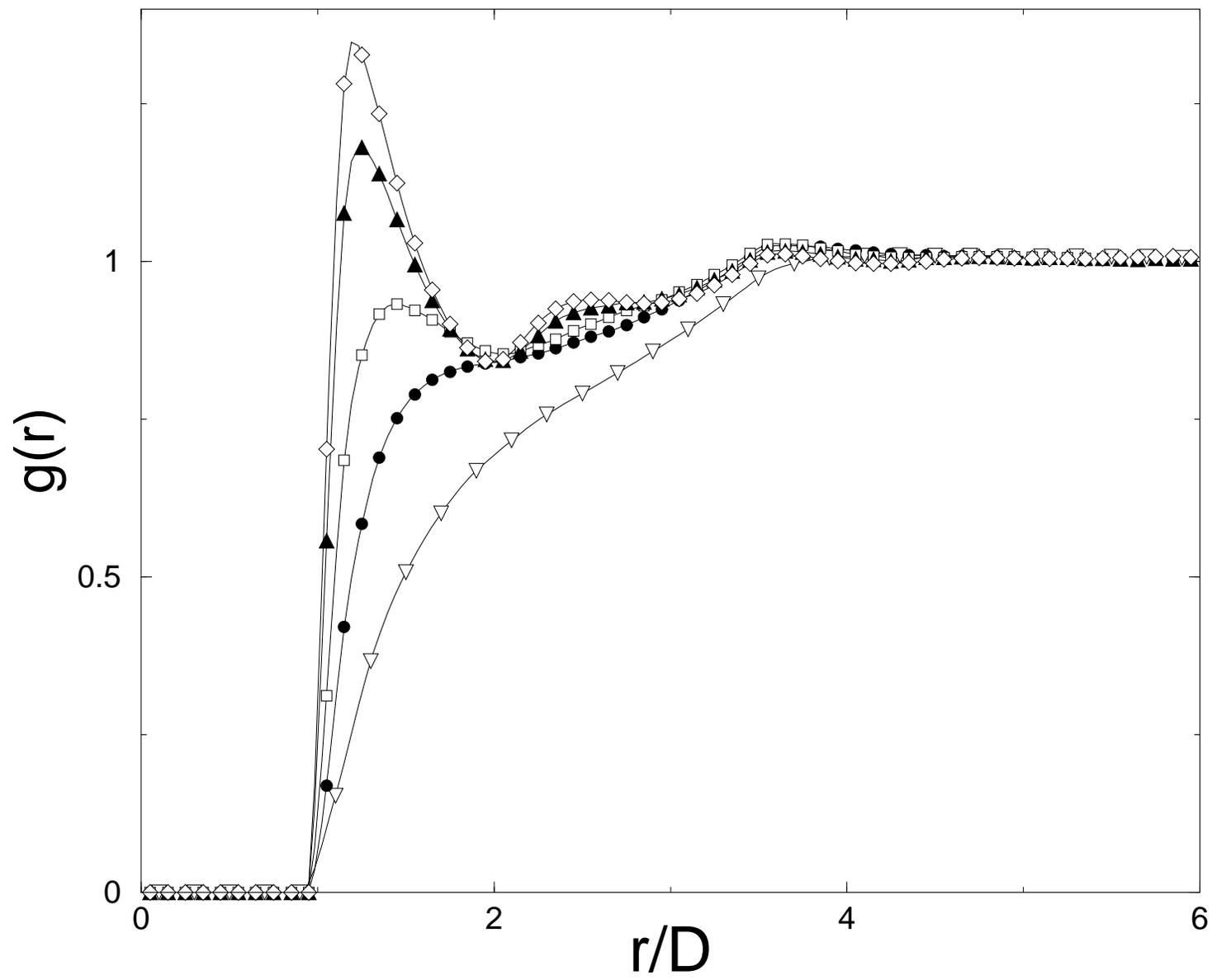
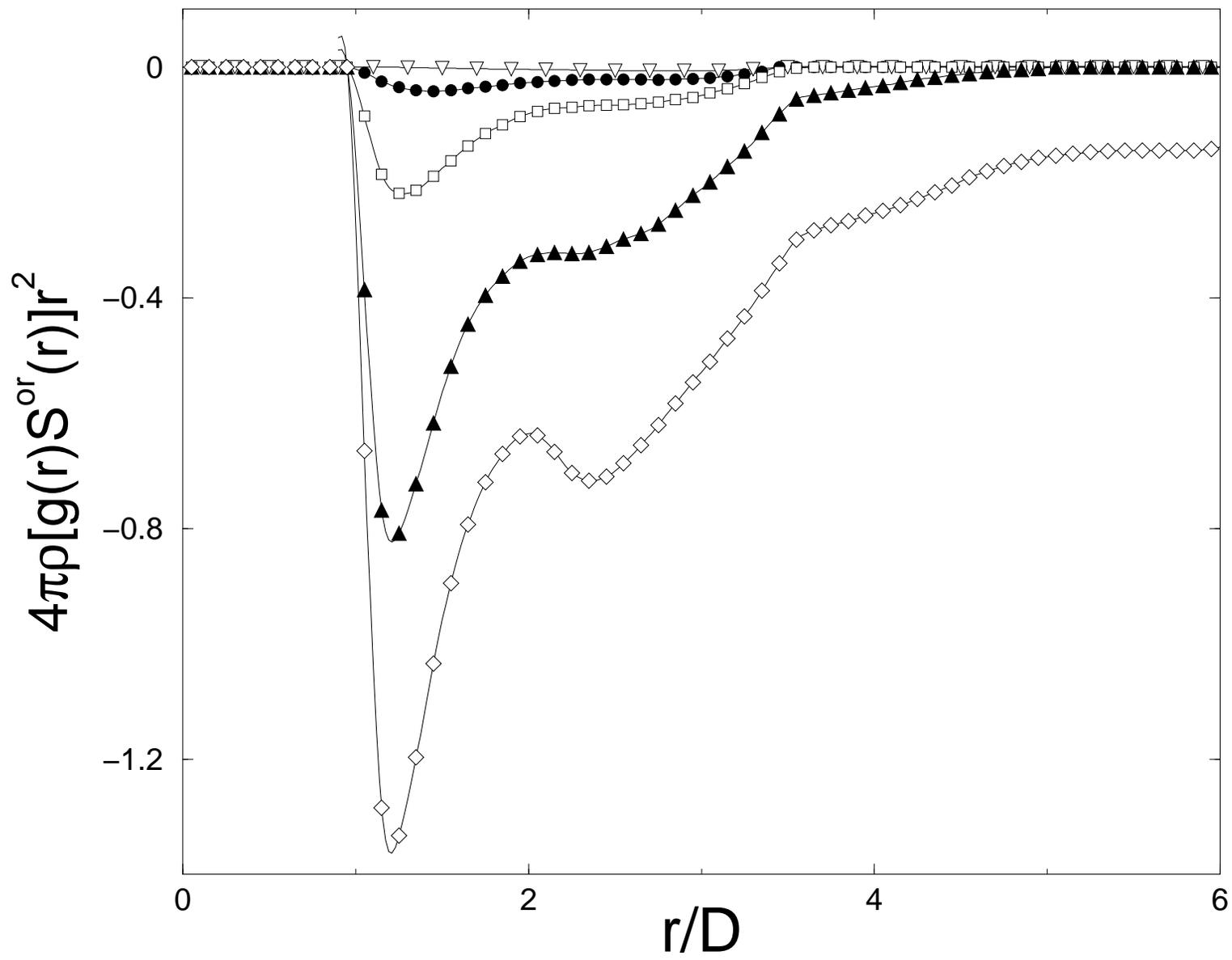
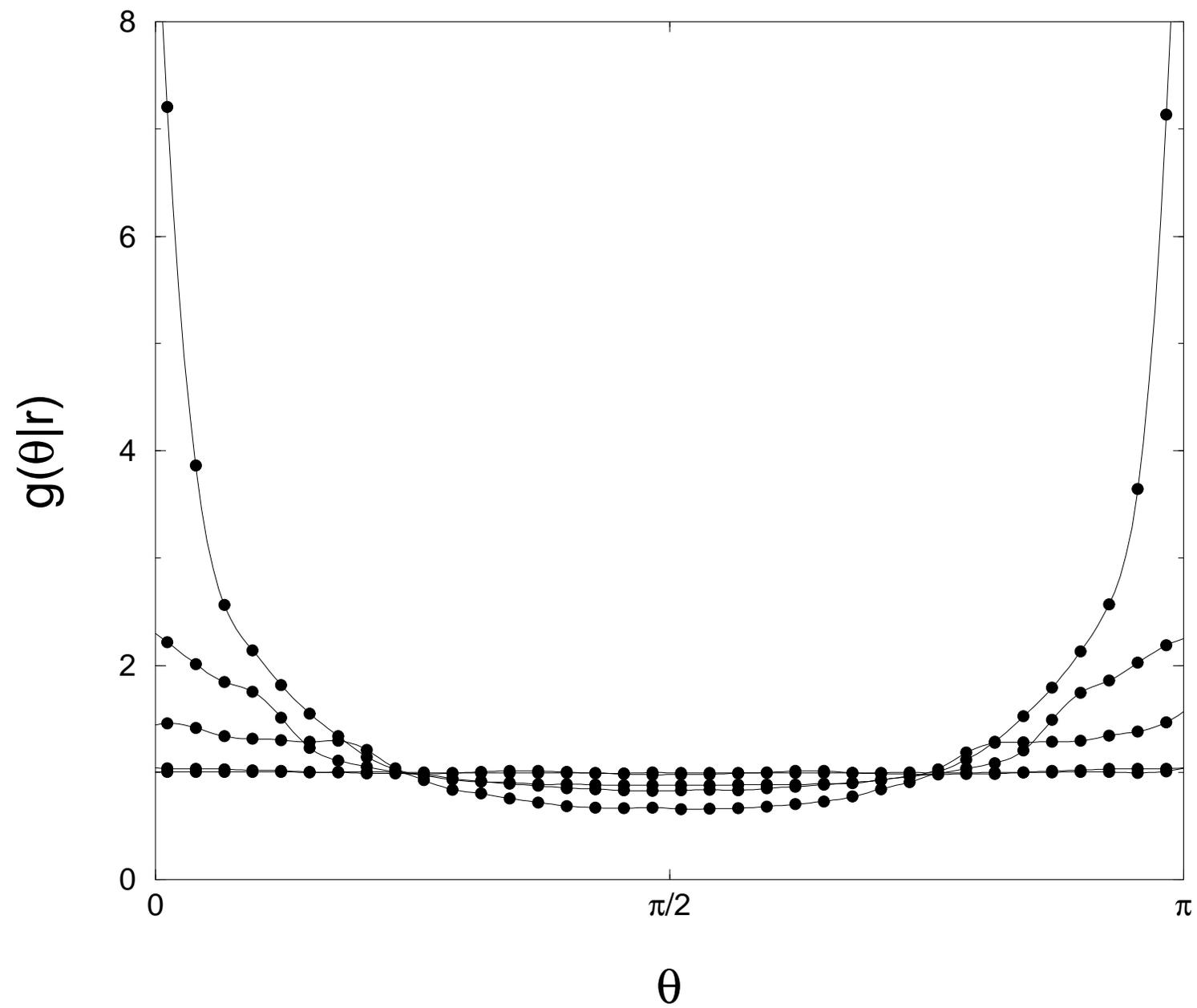
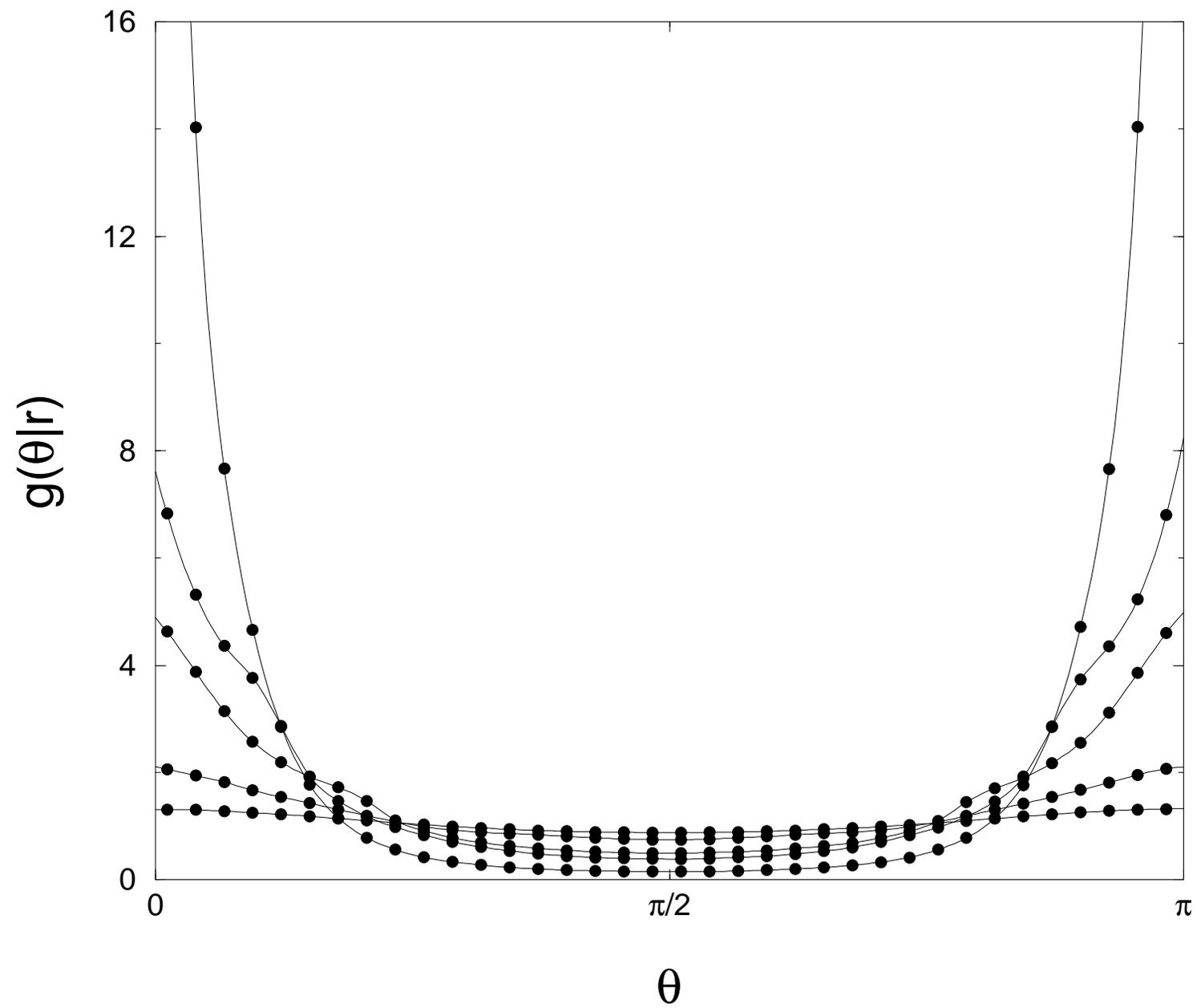
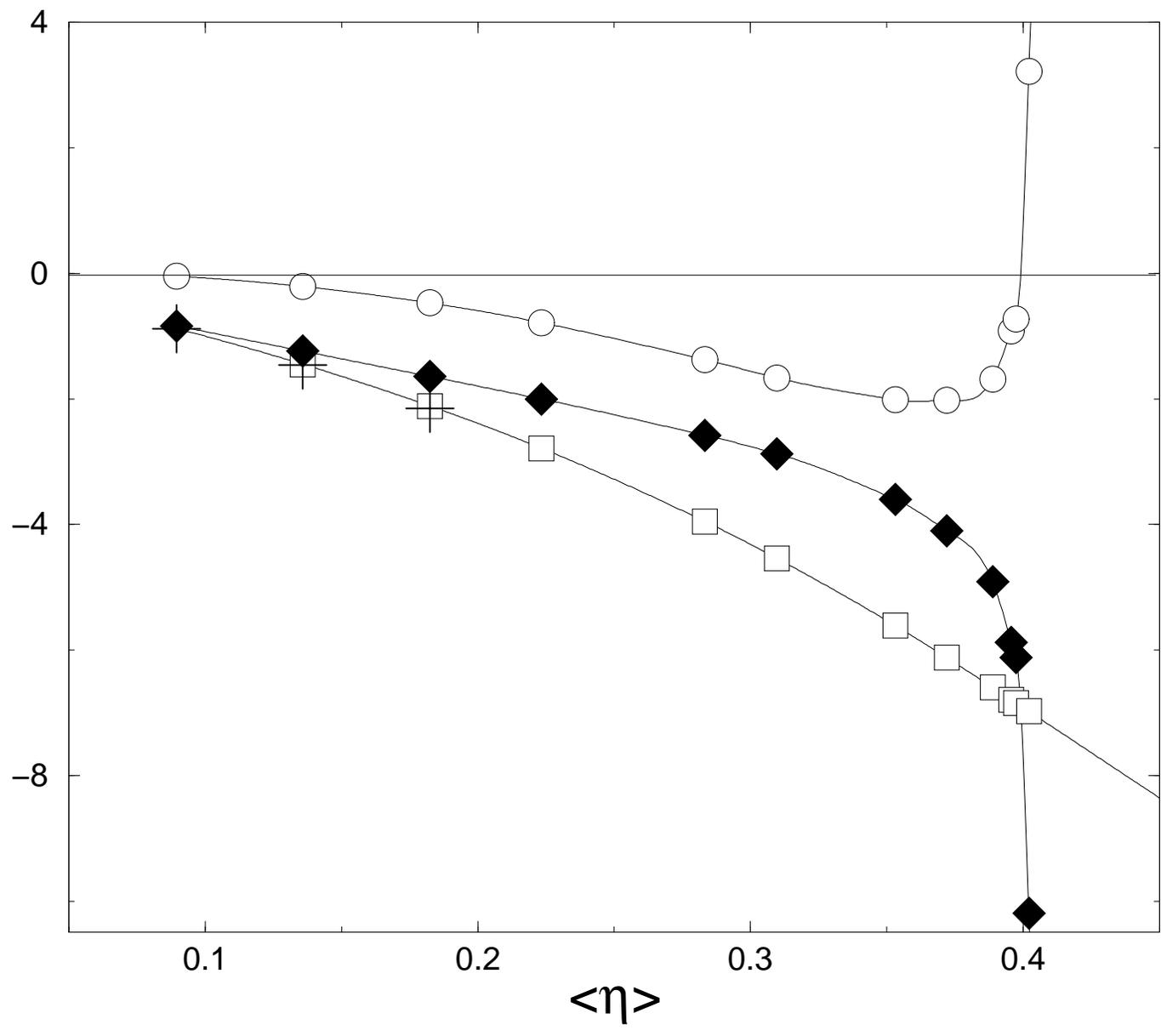
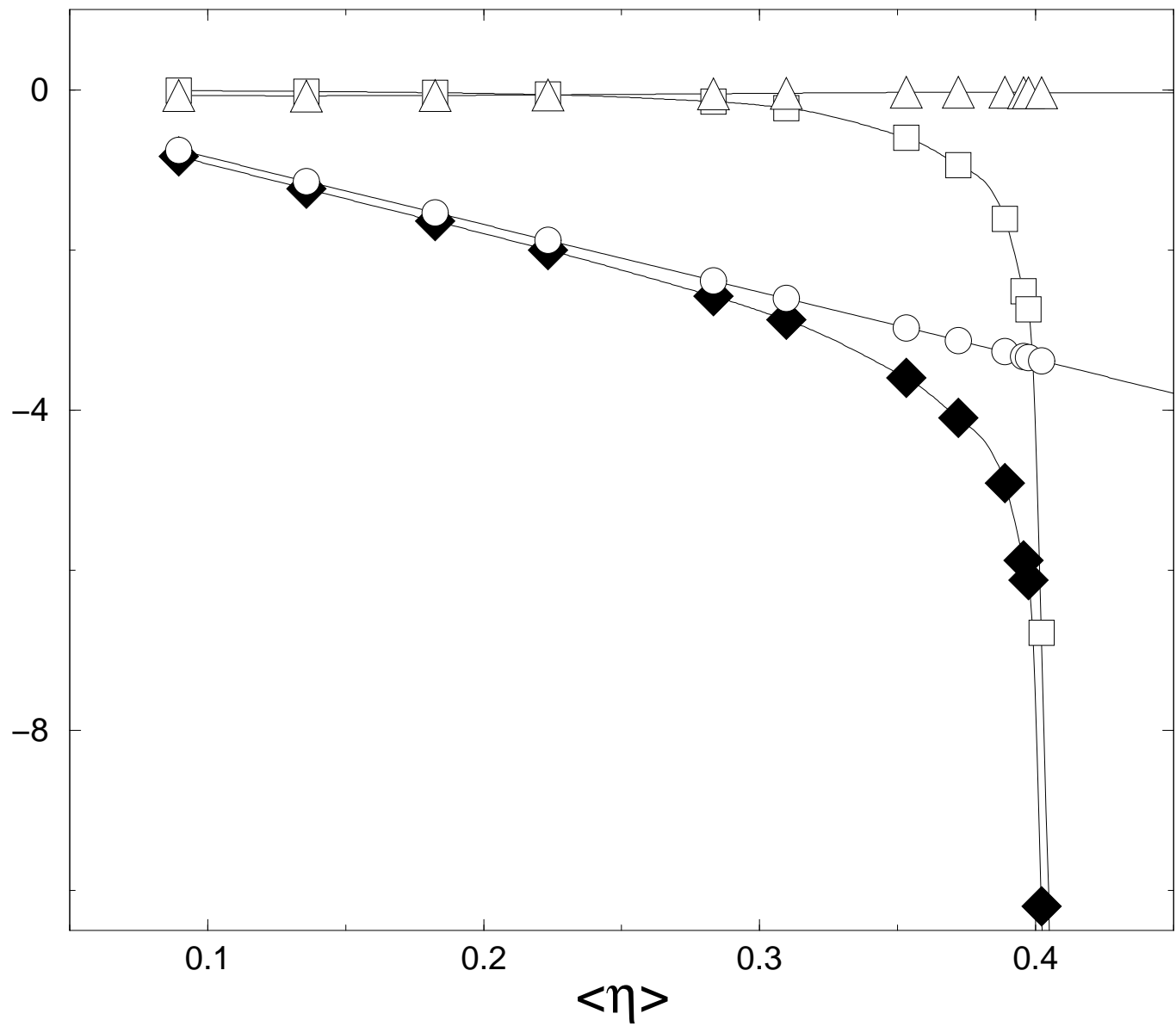
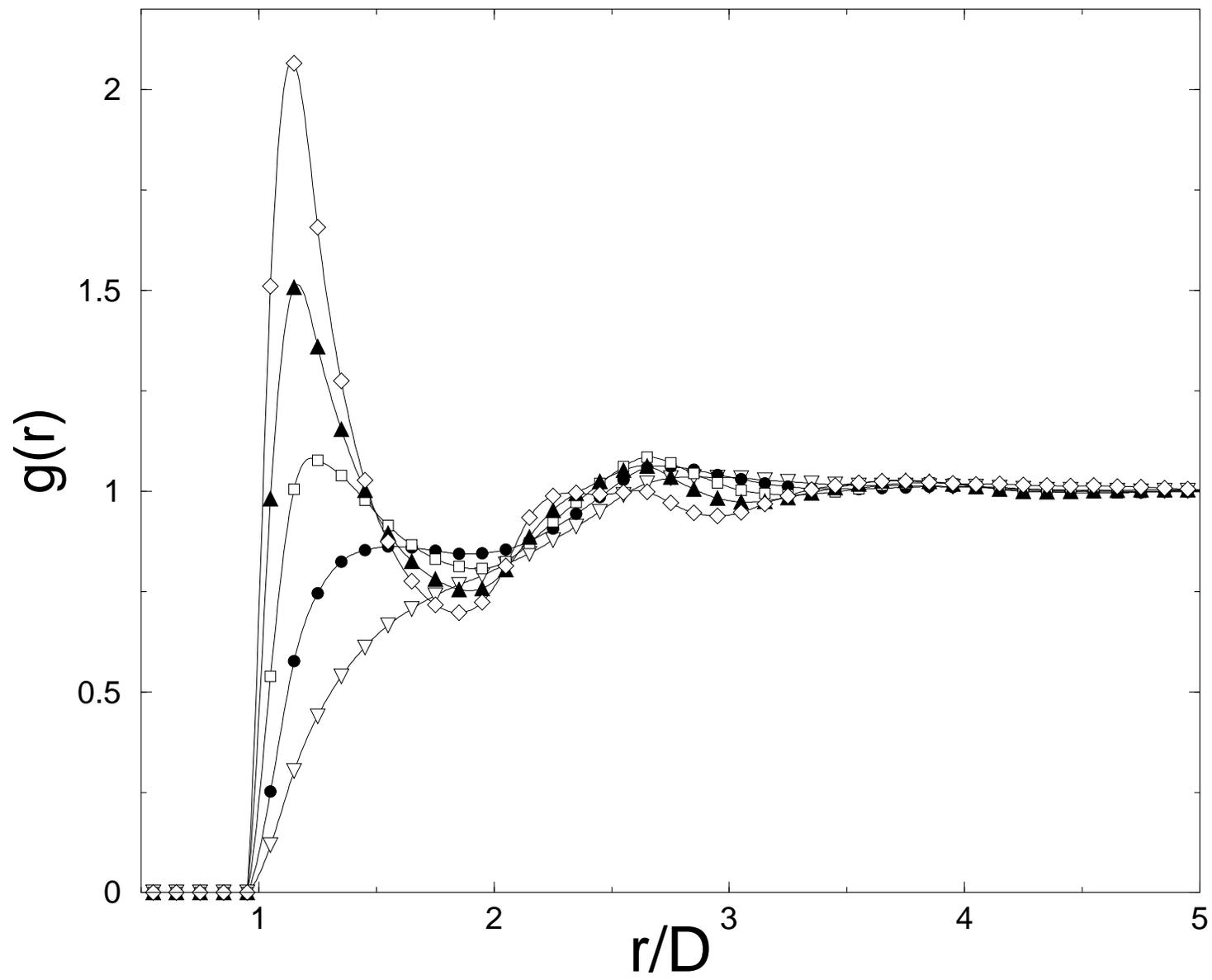
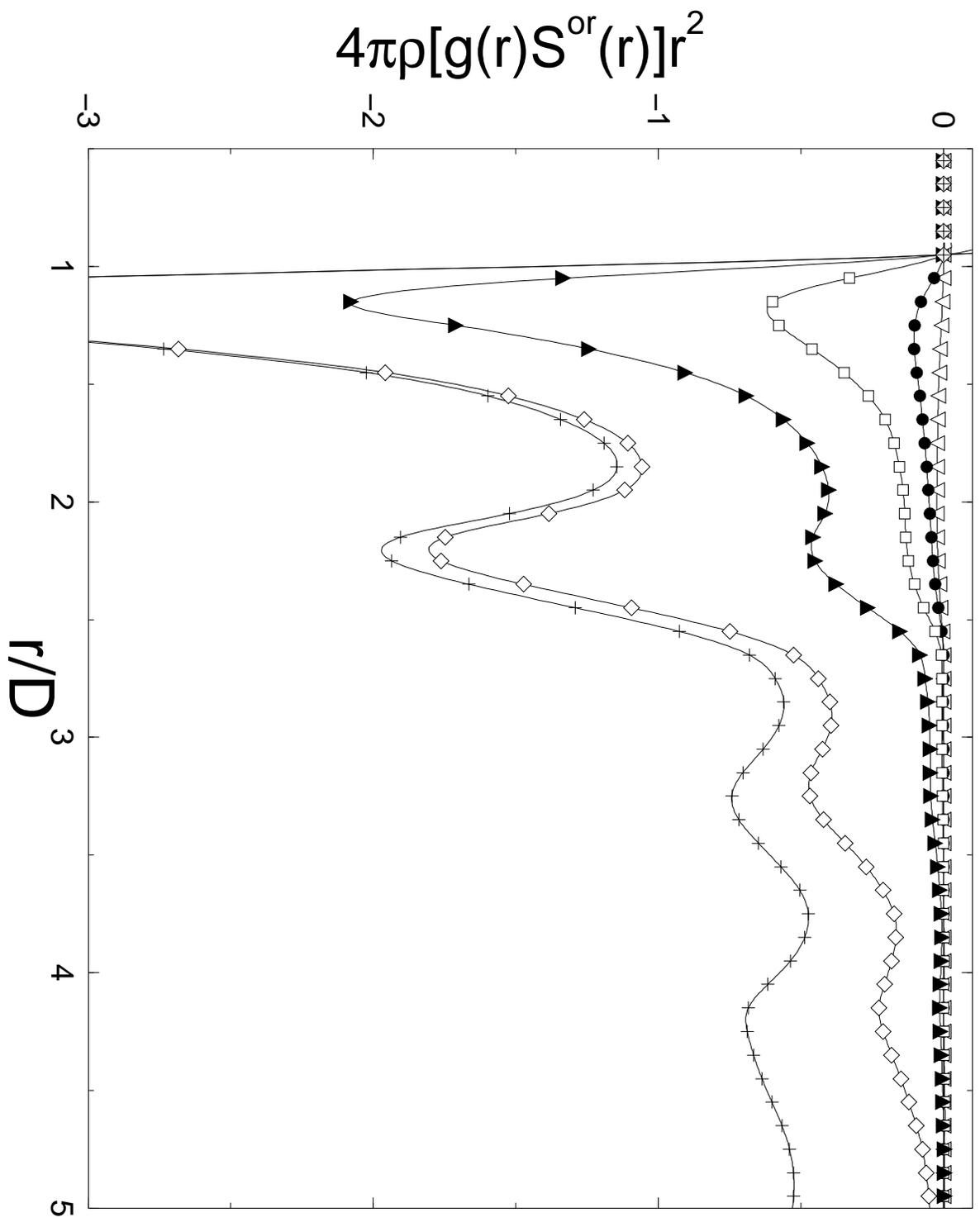
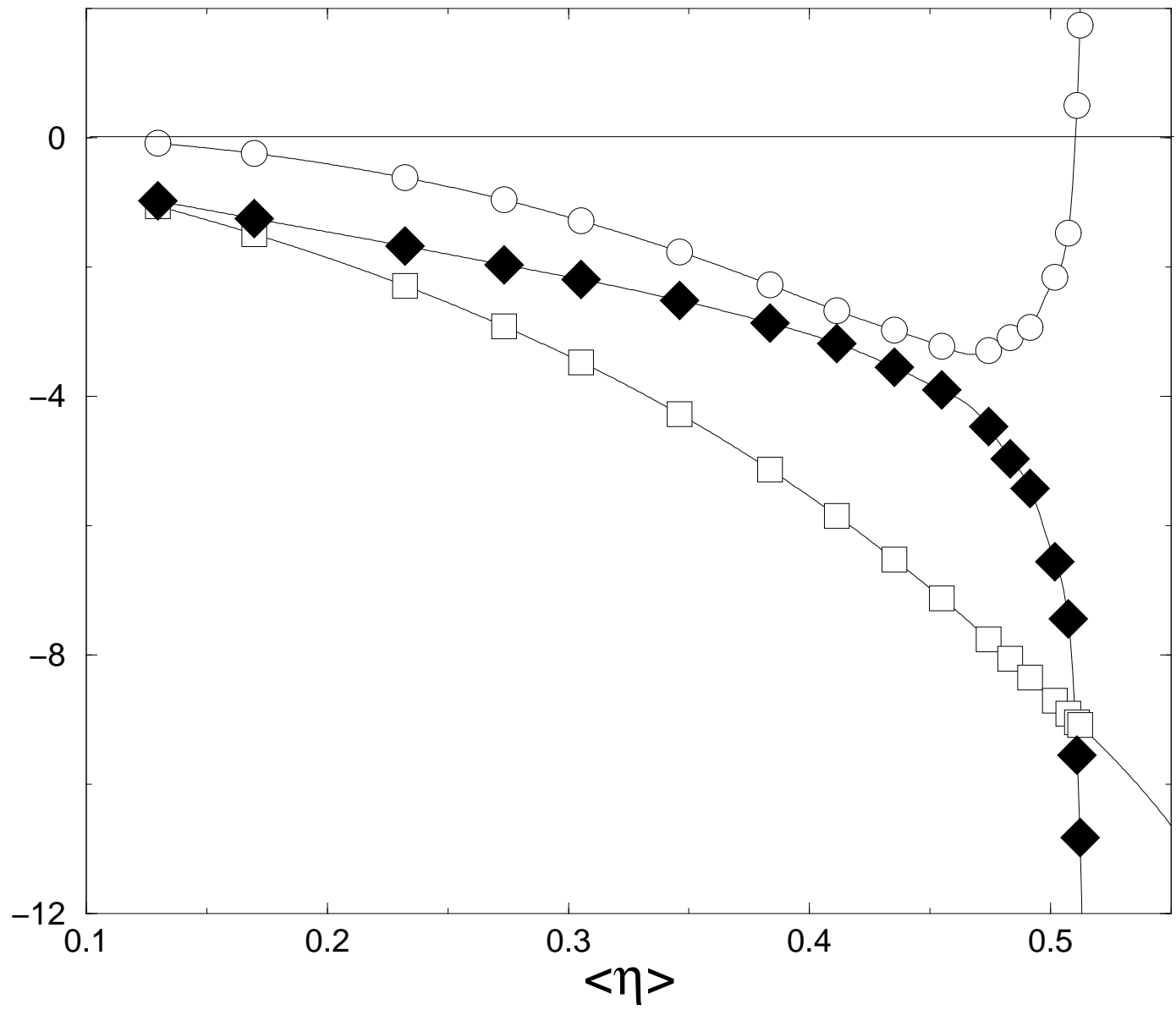
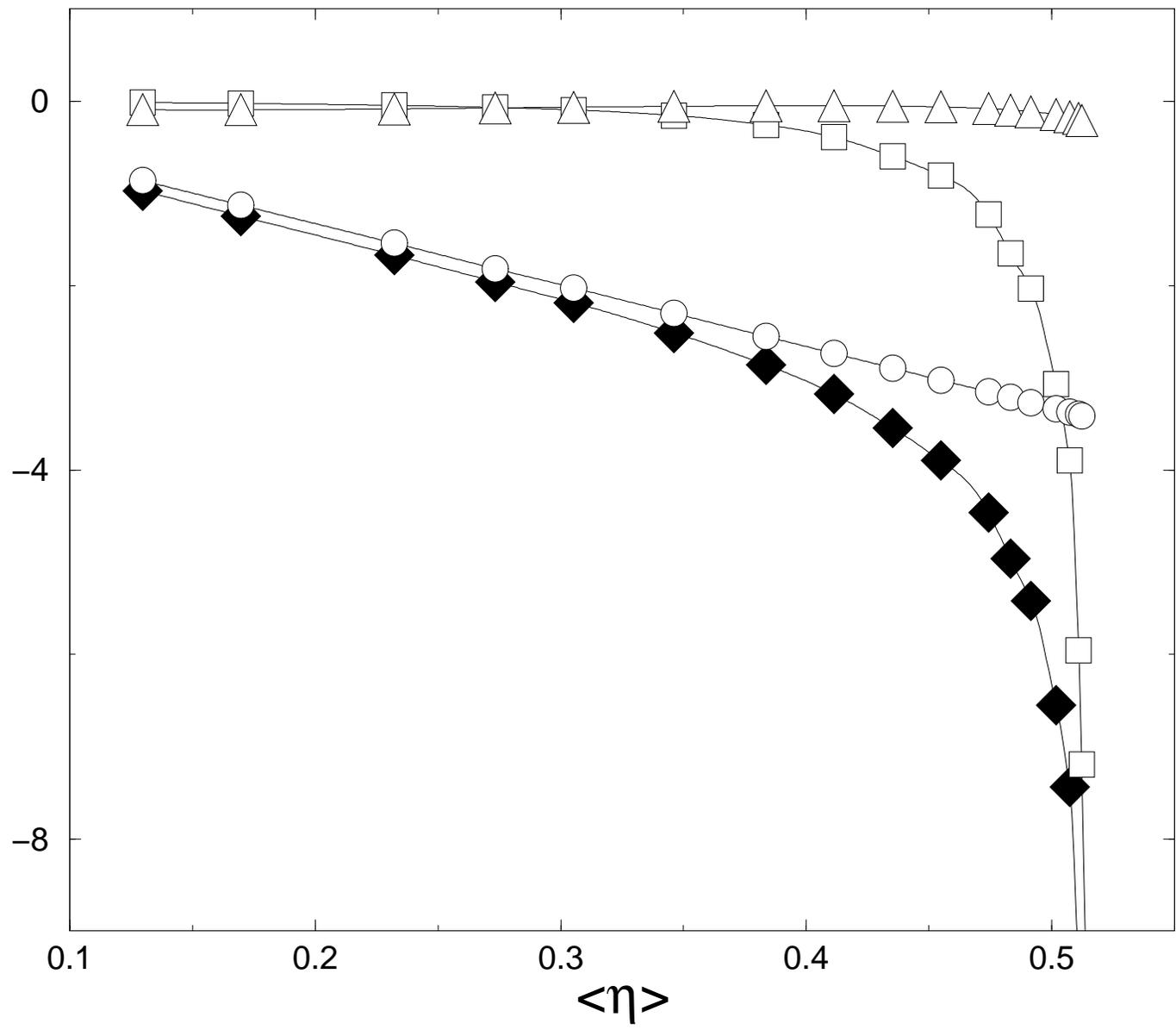
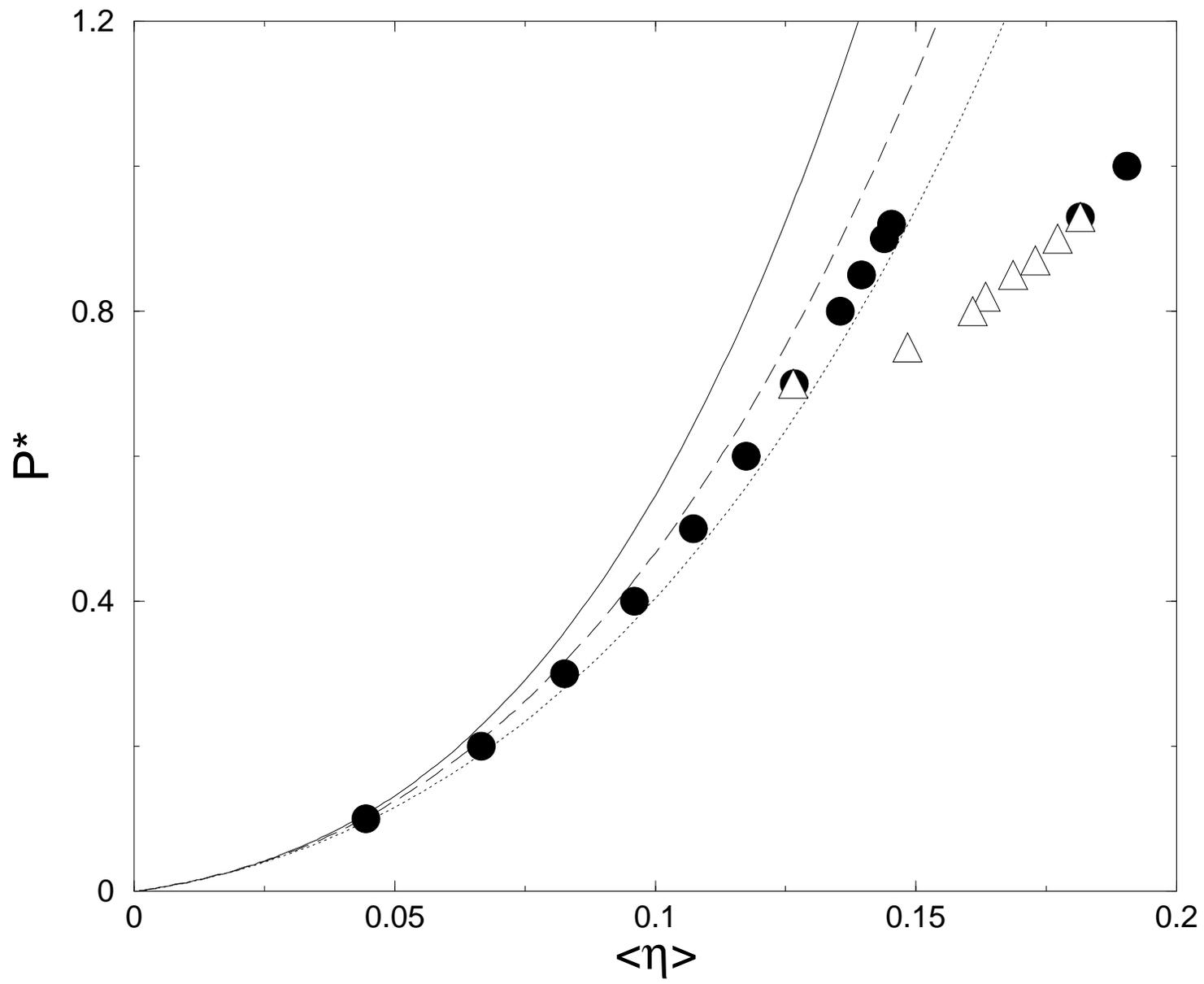
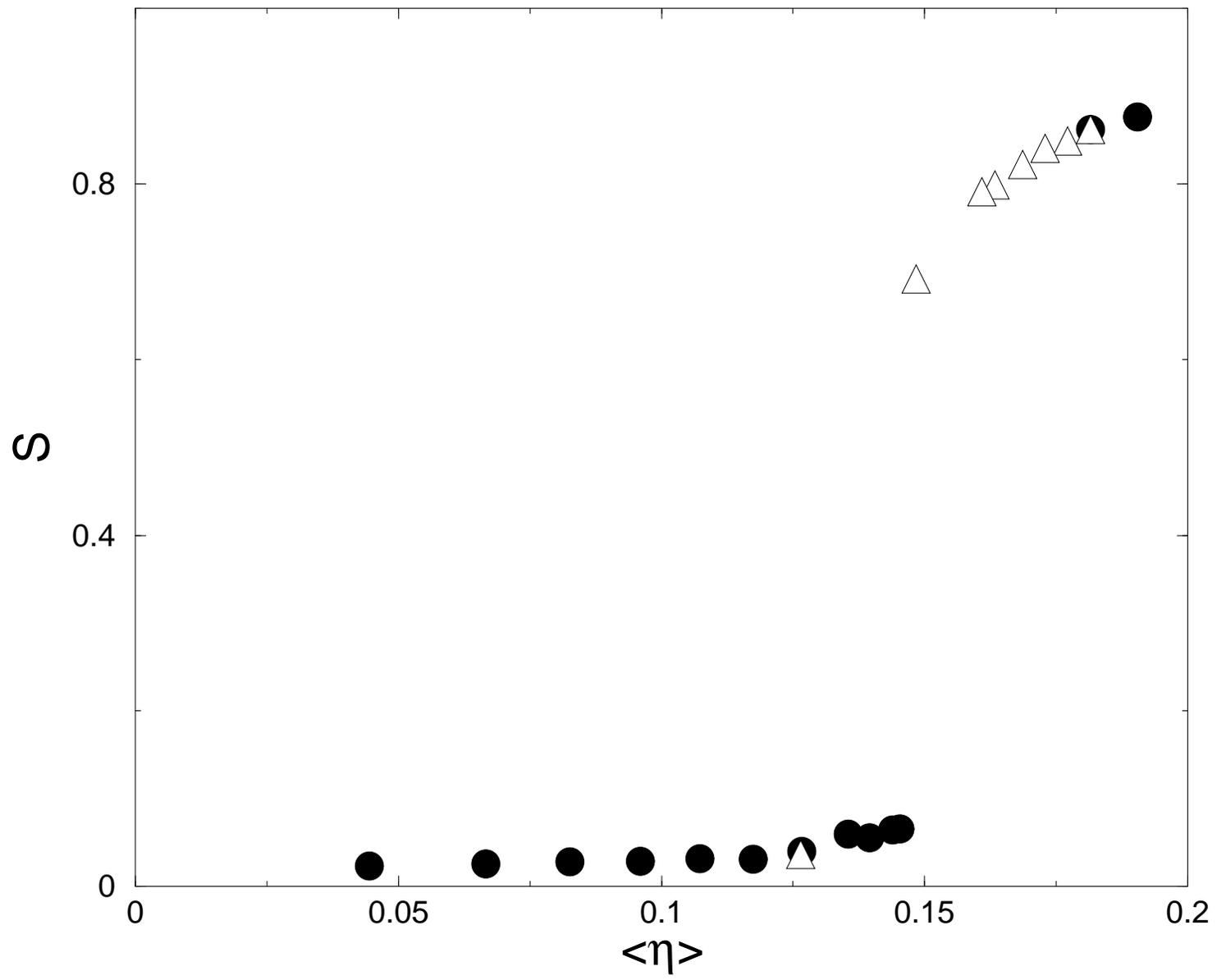
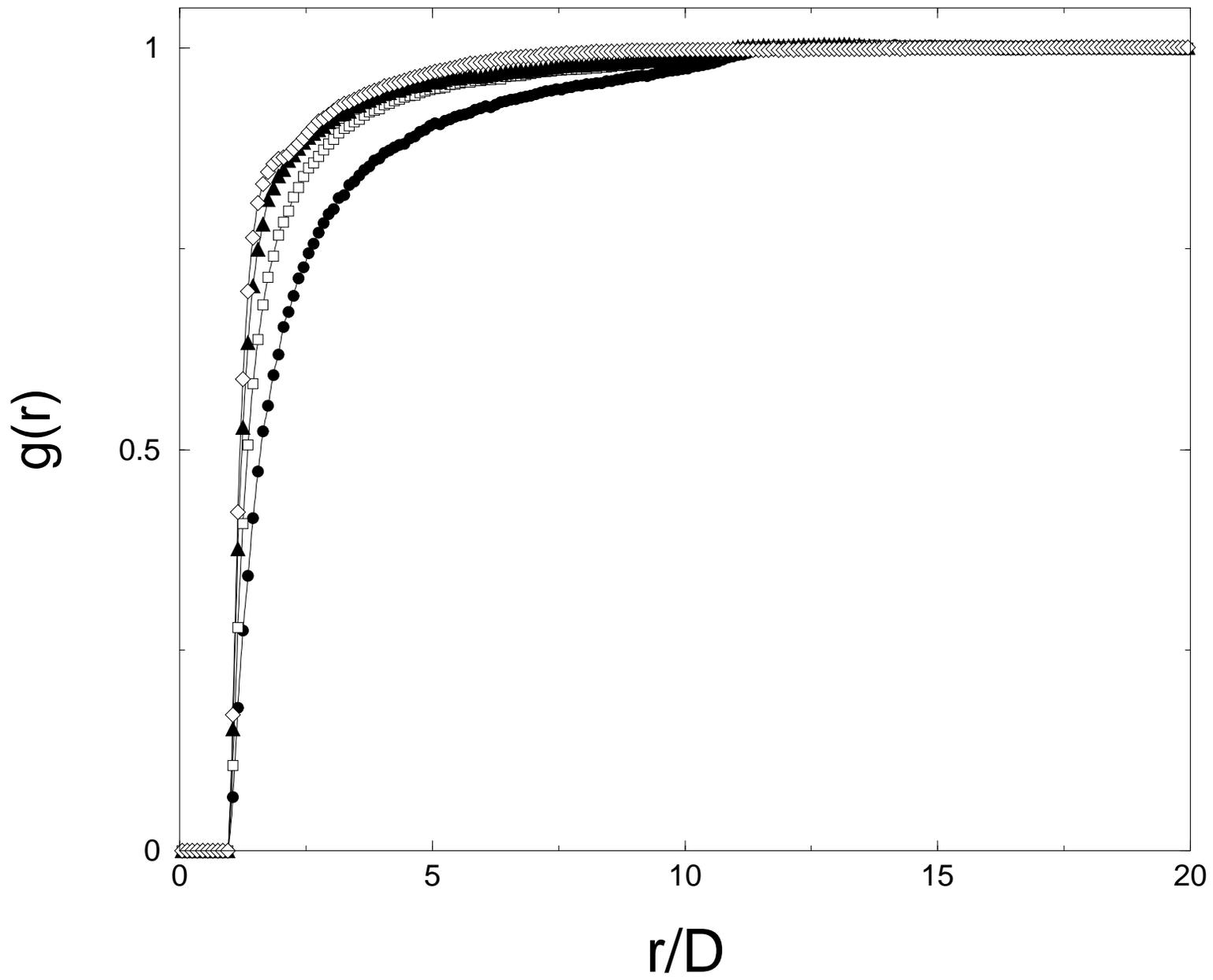
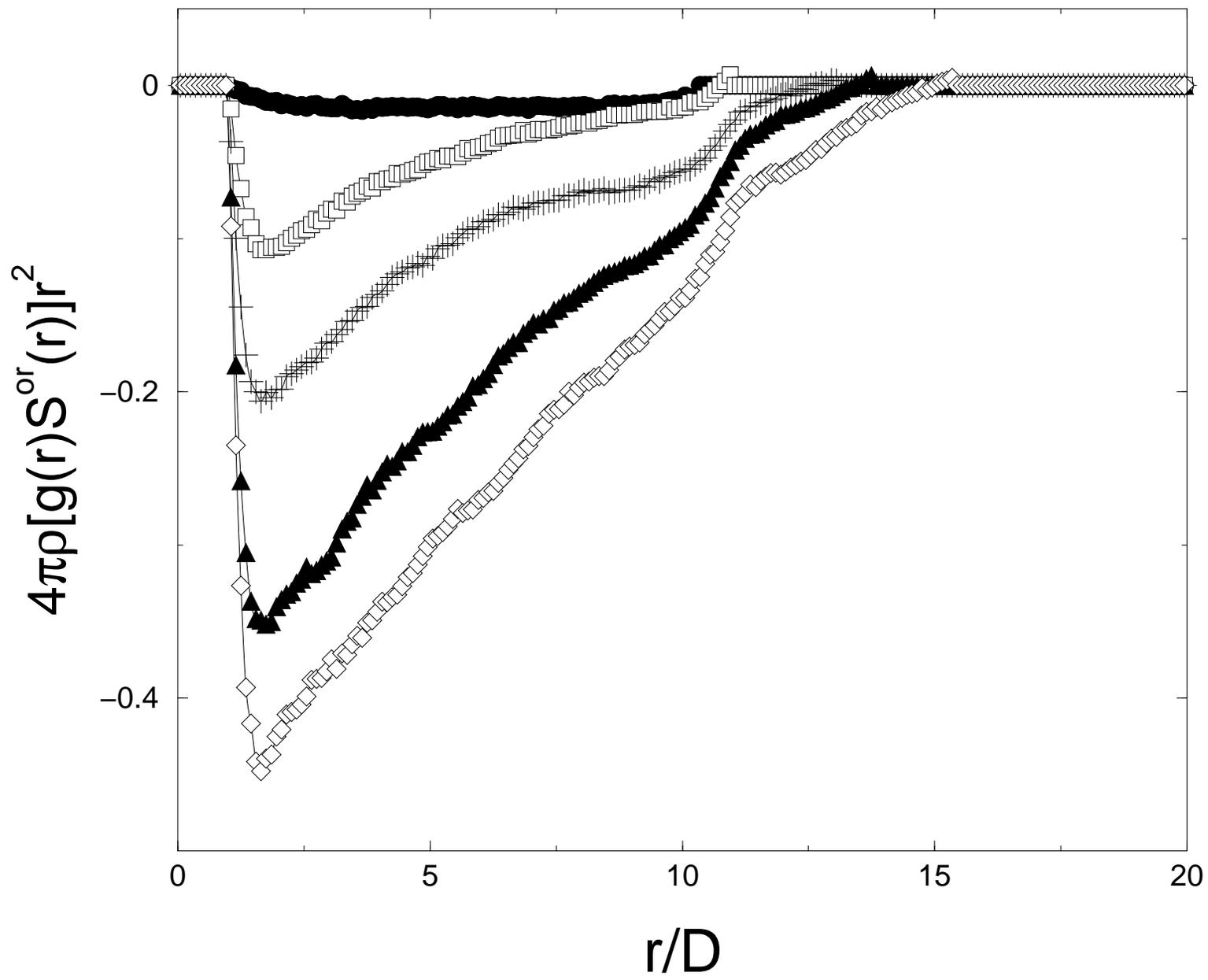
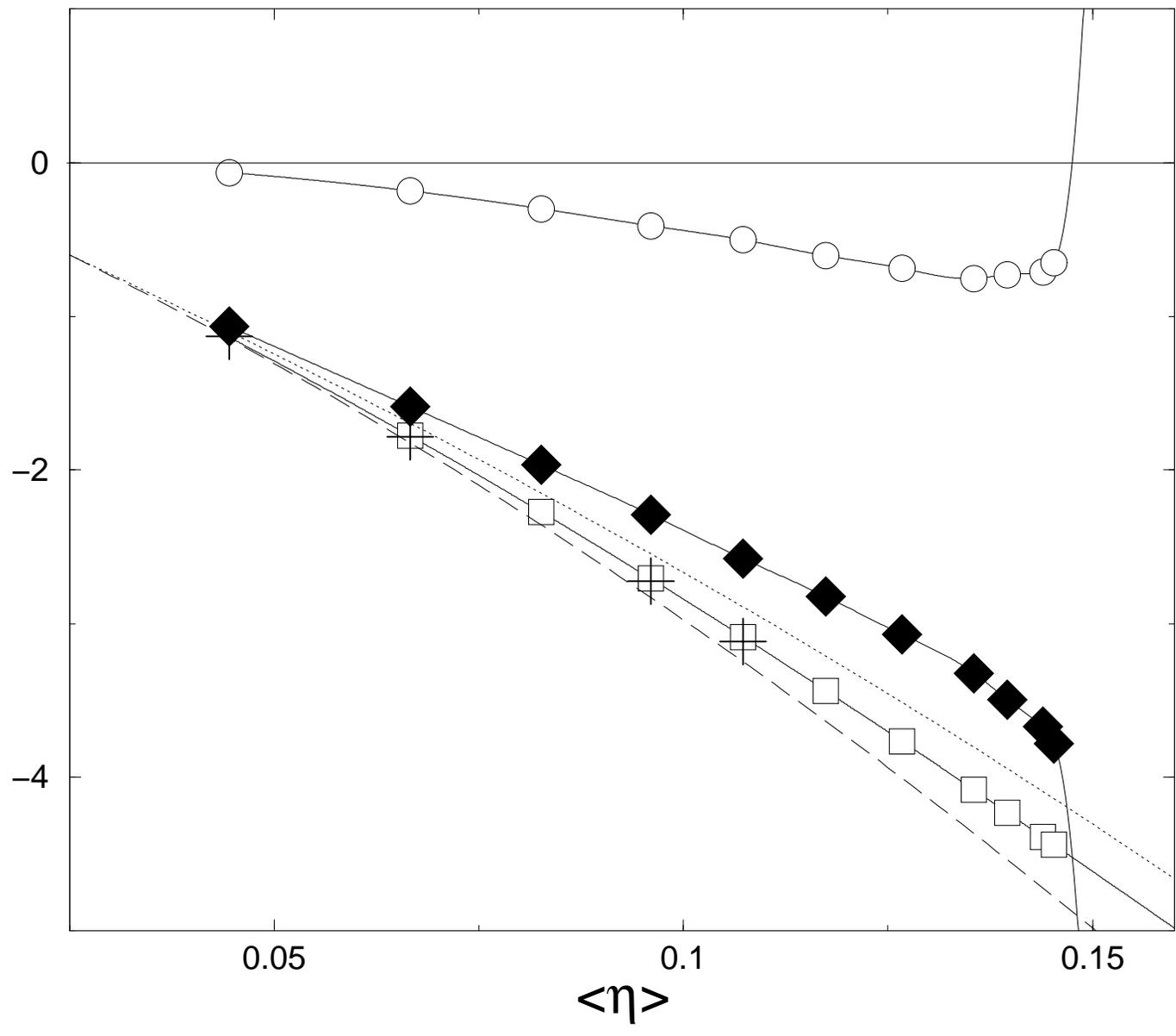
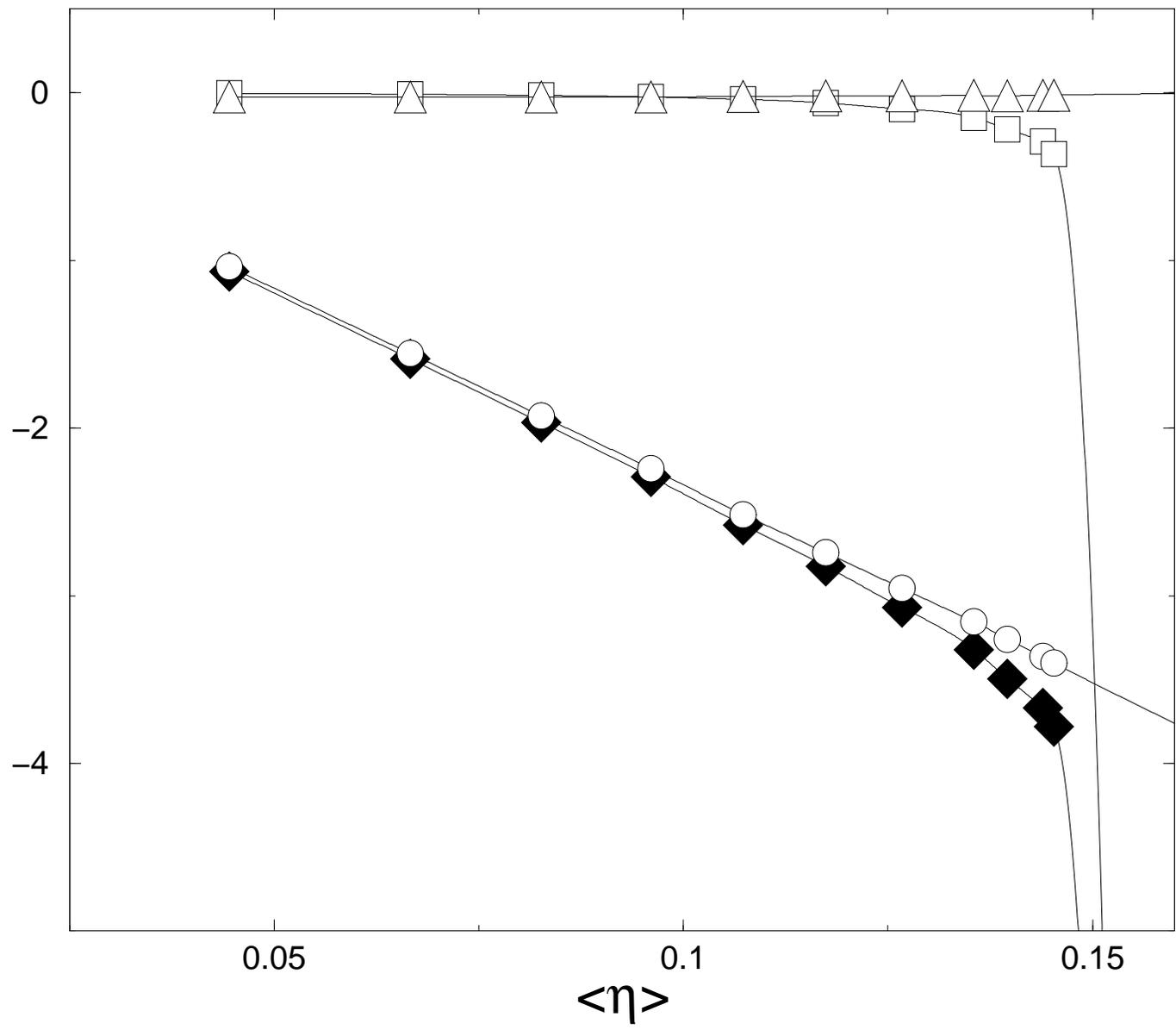
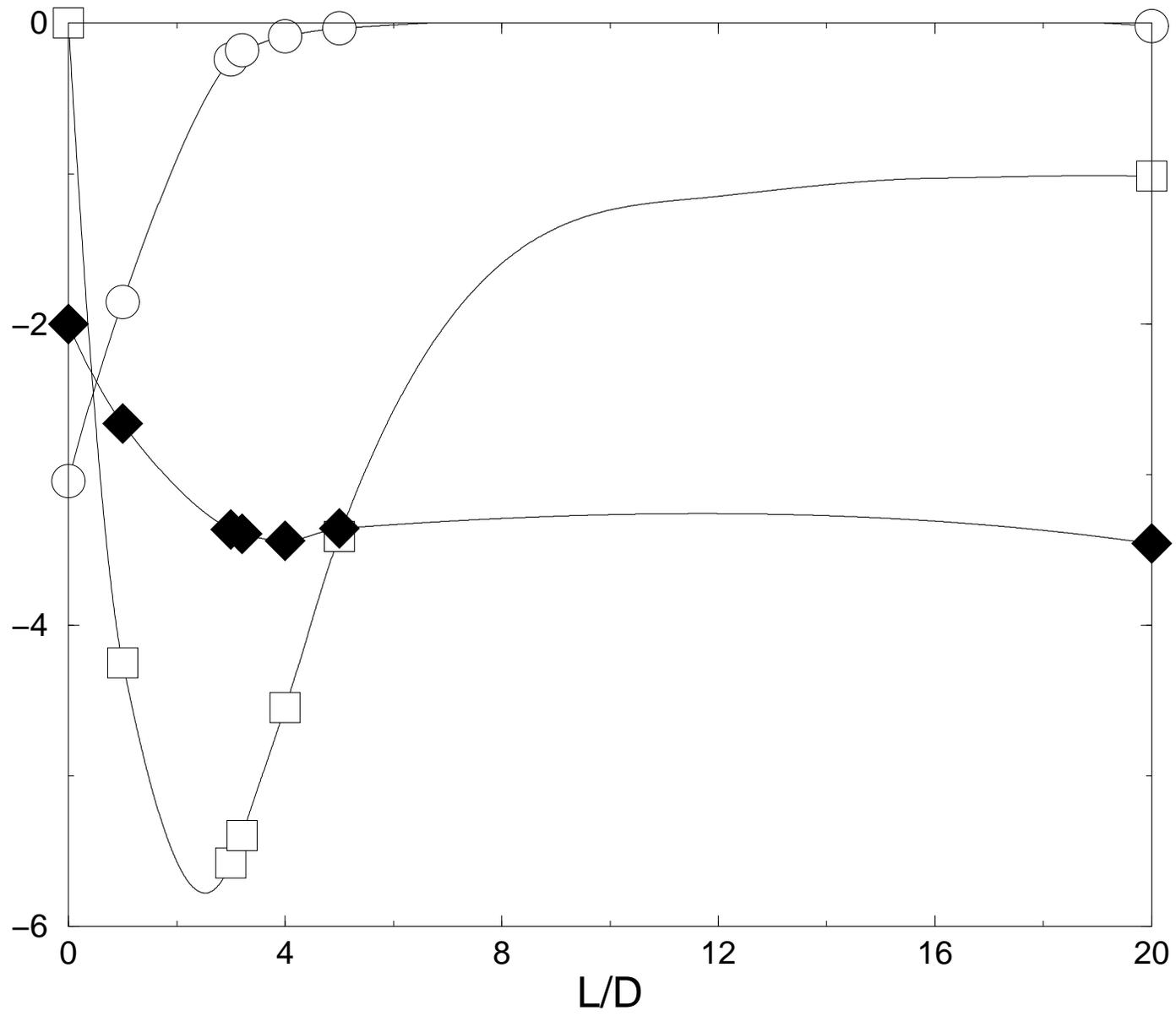
\end{document}